\documentclass[%
 reprint,
nofootinbib,%
 amsmath,amssymb,
 aps,
floatfix,
]{revtex4-2}

\usepackage{graphicx}
\usepackage{dcolumn}
\usepackage{bm}
\usepackage{multirow}
\usepackage{hyperref}

\hypersetup{
    colorlinks = true,    
    linkcolor = blue,     
    citecolor = blue,     
    urlcolor = blue,      
}
\begin{document}

\preprint{APS/123-QED}

\title{\textbf{\textbf{Testing Van der Waals black holes using black hole photon rings}}}

\author{
  Chengzhen Li$^{1}$,
  Meirong Tang$^{1}$,
  Zhaoyi Xu$^{1}$\thanks{Corresponding author: zyxu@gzu.edu.cn }
}
\email{zyxu@gzu.edu.cn(cncorresponding author) }

\affiliation{$^{1}$ College of Physics, Guizhou University, Guiyang, Guizhou 550025, China}

\date{\today}

\begin{abstract}
The Van der Waals black hole (BH) is a non‑Kerr model in Anti‑de Sitter (AdS) spacetime that exhibits characteristic thermodynamic phase transitions. Whether the physical validity of parameters such as the BH molecular volume $b$ and pressure $P$ is feasible still requires observational verification. As a sensitive probe of strong gravitational fields, the BH photon ring directly carries spacetime geometric information, making it an ideal object for model testing. In accordance with the spacetime metric associated with this BH, this study obtains the null geodesic equations for photons and conserved quantities, determines the photon sphere conditions and radii via effective potential analysis, investigates the photon ring classification and radiative properties, and employs measurements from the Event Horizon Telescope (EHT) to place constraints on parameters and simulate the optical appearance. The results show that the molecular volume parameter $b$ significantly regulates the radius of the photon sphere, photon ring structure, and the luminosity. Within a specific interval, the model exhibits a high degree of agreement with the EHT measurements for M87* and Sgr A*. This interval not only satisfies the energy conditions (weak, strong, and dominant) but also corresponds to a stable Van der Waals‑type phase transition region. Furthermore, the photon ring can distinguish between the Van der Waals BH and the Schwarzschild BH: due to the small‑black‑hole–large‑black‑hole (SBH–LBH) phase transition and microstructural discontinuities, the photon ring of the Van der Waals BH is particularly sensitive to the molecular volume parameter $b$, whereas the Schwarzschild BH lacks these features. This work uses the photon ring to probe BH molecular models, linking its microstructure, spacetime geometry and observational signatures, and provides a new way to derive BH microphysics from observations.

\end{abstract}

\maketitle
\section{Introduction}
\label{sec:level1}
BHs are extreme spacetime regions under strong gravitational backgrounds. Their spacetime geometry and thermodynamic properties not only serve as an effective arena to testing general relativity but also provide an important window to investigate quantum gravity effects \cite{ref1,ref2,ref3,ref4}. The microscopic molecular model for BHs, as constructed by Wei and Liu \cite{ref5}, furnishes a novel theoretical framework for comprehending the correlation between the microscopic physical traits of these compact objects and their macroscopic spacetime geometry. By incorporating thermodynamic phase transitions alongside statistical fluctuations, this theoretical framework elucidates the statistical characteristics inherent to the microscopic degrees of freedom of BHs, thereby opening novel avenues for the formulation and empirical validation of quantum gravity models \cite{ref6,ref7,ref8}. 

The Van der Waals BH, as an important non‑Kerr model in AdS spacetime, exhibits thermodynamic behavior closely resembling that of fluid systems. After introducing the BH molecular volume parameter and pressure parameter, the characteristic SBH–LBH phase transition emerges \cite{ref5,ref9}.Previous studies have established a link between the thermodynamics and geometrical optics of van der Waals BHs. Chabab et al. \cite{ref10} showed that the phase transition can be read off from the impact parameter and radius of unstable circular photon orbits, exhibiting a critical exponent of 1/2 that aligns precisely with the thermodynamic counterpart. In terms of quasinormal modes, Liu et al. \cite{ref11} discovered that, in the vicinity of the critical point, the slopes of the quasinormal frequencies of SBH and LBH with respect to parameters such as charge undergo abrupt changes. It has been established by Liang et al. \cite{ref12}, within the context of Lorentz-violating gravity, that the non-monotonic feature displayed by the imaginary part of the frequency emerges exclusively when the phase transition is underway. In addition, it has been demonstrated by Molla et al. \cite{ref13} that the influence exerted by Van der Waals parameters on both the BH shadow and strong gravitational lensing can be significantly magnified in the presence of plasma, thus facilitating their differentiation from conventional BH configurations.

The no-hair theorem tells us that a conventional BH has only three defining properties—mass, angular momentum, and charge—and nothing else is needed to specify it. Nevertheless, non-‐Kerr BHs predicted by quantum gravity theories, such as hairy BHs (e.g., scalar hairy BHs characterized by an inverted Higgs potential \cite{ref14}), Bumblebee BHs \cite{ref15}, and Horndeski gravity BHs \cite{ref16}, introduce additional parameters, thereby opening up new possibilities in BH physics. These parameters essentially reflect the microscopic structure of BHs or the interactions with surrounding matter fields. Their physical validity cannot be verified solely through theoretical derivations; instead, the plausibility of the models can only be confirmed by capturing the observable signatures associated with these parameters from observational data \cite{ref17}.

The optical manifestations of BHs, encompassing both shadow morphologies and photon ring structures, have been acquired by the EHT collaboration through its recent observations of M87* and Sgr A* \cite{ref18,ref19}. This has also supplied empirical data for testing BH models and opened new avenues for research in BH physics \cite{ref20,ref21,ref22,ref23,ref24,ref25,ref26}.

Arising from the superposition of photons that complete several revolutions along critical bound orbits—namely, the photon sphere—prior to their escape, the photon ring constitutes a distinctive higher-order optical signature within the intense gravitational field of a BH \cite{ref27,ref28,ref29,ref30,ref31,ref32}.The photon ring manifests as a delicate annular configuration, whose radius, luminosity profile, and the count of subring layers are essentially governed by the spacetime metric of the BH, while remaining largely insensitive to the specifics of the accretion disk model \cite{ref33,ref34,ref35}. In terms of specific classification, we divide the formation of the photon ring into two categories of key photon motions: one category consists of photons that escape after stably circling on the photon sphere orbits \cite{ref28}, whereas the remaining class is composed of photons that arrive at the observer subsequent to undergoing repeated bending induced by the BH's gravitational lensing \cite{ref36}. The constructive interference of such photons manifests the characteristic "narrow luminous annulus" of the photon ring \cite{ref27}.Serving as a pivotal link that bridges the macroscopic spacetime geometry and the microscopic physical attributes of BHs, the photon ring displays a markedly stronger dependence on BH parameters relative to the shadow \cite{ref37,ref38}. For instance, its radius is closely related to the spacetime curvature distribution governed by the BH microstructure \cite{ref39}, while its brightness peak reflects the combined effects of photon orbital structure and radiative transfer in strong gravitational fields \cite{ref34}. This characteristic makes the photon ring an ideal probe for distinguishing Kerr BHs from non‑Kerr ones and for constraining model parameters \cite{ref27,ref40,ref41,ref42}.

Recent high‑precision EHT observations have released the morphological data of the M87* photon ring, providing stronger constraints for tests of non‑Kerr spacetimes \cite{ref43}. It should be noted, however, that non‑Kerr models suffer from parameter degeneracy issues, making it difficult to distinguish them accurately solely by shadow data. The fine structure of the photon ring resolves this difficulty \cite{ref40}, which further underscores the importance of this study focusing on the photon ring characteristics of Van der Waals BHs.

Previous studies have demonstrated that, in comparison with the BH shadow, both the photon ring and the photon sphere—the former referring to the superposition of photons that escape from the photon sphere and the latter to the critical bound orbits of photons near the BH—possess a heightened sensitivity to the underlying parameters. This feature makes them effective discriminators among different BH models \cite{ref37,ref38,ref44}. The employment of photon rings as a means to probe gravitational theories has attracted considerable attention in recent years, encompassing investigations into the optical appearance of BHs within Horndeski gravity \cite{ref45} as well as regular BHs in the asymptotically safe gravity framework \cite{ref46}. These studies collectively furnish valuable guidance for the present work, in which photon rings are utilized to examine van der Waals BHs \cite{ref40,ref44}. Meanwhile, investigations into the microstructure of AdS BHs have indicated that photon motion is related to the weak gravitational interactions among BH molecules, thereby providing theoretical support for analyzing the photon rings of Van der Waals BHs \cite{ref10,ref47,ref48}.

Furthermore, thermodynamic studies \cite{ref5,ref9,ref49,ref50,ref51} have shown that the phase transition between SBH and LBH in charged AdS BHs belongs to the Van der Waals type. By identifying the pressure $P$ with the cosmological constant $\Lambda$ through $P = -\dfrac{\Lambda}{8\pi}$, one can match the BH to the Van der Waals fluid in a complete thermodynamic sense. For this matching, the equation of state and the critical parameters have already been computed precisely.

Following the framework of articles \cite{ref5,ref52}, this model introduces the BH molecular specific volume $v$ , establishing a quantitative connection between macroscopic thermodynamic quantities (such as pressure $P$ and volume $V$) and microscopic degrees of freedom. Such a micro-macro correspondence leaves its imprint on the spacetime geometry via corrections to the metric function $f(r)$, thereby rendering it accessible to photon-ring observations. The Ruppeiner geometry formalism provides a framework in which the scalar curvature of charged AdS BHs exhibits a sign change. Specifically, it is positive for small BHs and negative for large ones, which signals repulsive and attractive intermolecular interactions, respectively. At the critical point, both the sign reversal and the fact that the scalar curvature diverges closely parallel the behaviour found in a Van der Waals fluid. Hu et al. \cite{ref52} further systematized the geometric thermodynamic approach by embedding phase transition information into the geometric structure of phase space. This is essentially an effective theory: without presupposing a microscopic model of quantum gravity, it enables an effective description of the microstructure to be inferred solely from macroscopic phase transition behavior, thereby providing a new entry point for the microscopic foundation of BH statistical mechanics. These findings directly explain why the molecular volume parameter $b$ can regulate photon motion—the variation of $b$ is actually the macroscopic manifestation of the BH microscopic molecular volume. It modifies the spacetime curvature by altering the molecular distribution and interactions within the BH, which in turn affects photon orbits and the characteristics of the photon ring.

Building on what has been covered above, the remainder of this paper is laid out in the following way. Sect. II presents the spacetime metric of the Van der Waals BH and its connection with the BH molecular number density, thereby establishing the geometric and thermodynamic basis for the subsequent analysis of photon motion. In Sect. III, we first derive the null geodesic equations in a systematic way, and then use the effective potential to determine the conditions and radius of the photon sphere. The radiative contributions from photon rings are classified, and the model parameters are constrained by employing the observational measurements reported by the EHT. In Sect. IV, the optical appearances of the BH are simulated across various emission models, and we then quantify how the photon ring radius, its brightness, and the number of ring layers respond to the molecular volume parameter $b$. In Sect. V, the primary findings are summarized, along with a discourse on their physical implications. Through the above theoretical derivations and observational simulations, this paper aims to systematically elucidate the physical mechanism by which the Van der Waals BH parameter $b$ modulates the photon ring structure, and to verify the consistency of the model with EHT data, thereby providing new insights for the observational test of this model and for the study of BH microphysics.

Throughout this work we work in the geometric unit system unless stated otherwise: $c=G=1$ and the BH mass is set to $M=1$. The metric signature is taken to be $(-,+,+,+)$.

\section{VAN DER WAALS BH METRIC AND NUMBER DENSITY}
\label{sec:level2}
The thermodynamic properties of BHs originate from their spacetime geometry. For the BH's thermodynamic properties to exhibit a precise correspondence with the corresponding properties of the Van der Waals fluid, an asymptotically AdS metric must be appropriately formulated. Under the premise of static spherical symmetry, the line element for the Van der Waals BH is taken to be \cite{ref5} 
\begin{equation}
ds^{2}=-f(r)\,dt^{2}+\frac{dr^{2}}{f(r)}+r^{2}\,d\Omega^{2},
\label{eq:1}
\end{equation}
in which $d\Omega^{2}=d\theta^{2}+\sin^{2}\theta\,d\phi^{2}$ represents the line element on the two-dimensional sphere, while the metric function $f(r)$ acts as the pivotal entity that governs both the spacetime geometry and the thermodynamic properties. The line element is constructed by imposing the Einstein field equations, $G_{ab} + \Lambda g_{ab} = 8\pi T_{ab}$, in which $G_{ab}$, $g_{ab}$, and $T_{ab}$ denote the Einstein tensor, the metric tensor, and the stress-energy tensor, respectively. Upon merging the Van der Waals fluid equation of state with the first law of BH thermodynamics, and introducing the ansatz $h(r, P) = A(r) - P B(r)$ to resolve the corresponding partial differential equations, setting the integration constant $C_1 = \frac{8\pi}{3}$ to preserve the AdS spacetime structure, and taking $r_0 = 2b$ to simplify the logarithmic terms, the exact form of the metric function is derived as~\cite{ref53}
\begin{align}
f(r) &= 2\pi a - \frac{2M}{r} + \frac{r^2}{l^2} \left( 1 + \frac{3b}{2r} \right) - \frac{3\pi ab^2}{r(2r + 3b)} 
\notag \\
&\quad- \frac{4\pi ab}{r} \log\left( \frac{r}{b} + \frac{3}{2} \right).
\label{eq:2}
\end{align}
Under the approximation $b \ll r$ (i.e., the BH scale is much larger than the BH molecular volume), the metric function reduces to
\begin{align}
f(r) &= 2\pi a - \frac{2M}{r} + \frac{8\pi P}{3} r^2 \left( 1 + \frac{3b}{2r} \right) 
\notag \\
&\quad- \frac{4\pi a b}{r} \log\left( \frac{r}{b}\right)  - \frac{15\pi a b^2}{2r^2} + O[\left( \frac{b}{r} \right)^3],
\label{eq:3}
\end{align}
This approximate form highlights the dominant influence of the thermodynamic pressure $P$ on the spacetime geometry, which is consistent with the core features of the AdS spacetime metric, while retaining the key correction terms associated with BH molecular interactions. It thus provides a concise yet precise geometric foundation for subsequent investigations of photon ring properties in conjunction with thermodynamic parameters. Regarding the choice of $P$, according to Ref.~\cite{ref53}, the energy constraints are related to the value of $P$: the smaller the value, the broader the allowed range. With the aim of preserving proximity to the pure Schwarzschild BH scenario, the parameter is fixed at $P = 1 \times 10^{-6}$. This choice, on the one hand, satisfies the energy condition constraints required by Ref.~\cite{ref53}; conversely, this guarantees that the departure of the BH spacetime geometry from the Schwarzschild solution remains confined to the order of $10^{-6}$, thereby guaranteeing that the subsequent analysis of photon ring properties is both general enough and readily comparable with the Schwarzschild BH results. Moreover, this numerical value can be precisely represented in double‑precision floating‑point arithmetic, which facilitates numerical implementation.

Subsequently, the metric parameters $a$ and $b$ are assigned specific values. Within the framework of the Van der Waals fluid, the parameter $a$ characterizes the magnitude of the attractive interaction among BH molecules. For the sake of generality, $a = \dfrac{1}{2\pi}$ is adopted, and the present work restricts itself to the following concise discussion without delving further. The physical significance of $a$ is not limited to the intermolecular attraction of the BH; its value is directly related to the topological structure of the BH horizon and the constant curvature of the two-dimensional surface~\cite{ref52}: when $a = \dfrac{1}{2\pi}$, the horizon is spherically symmetric; when $a = 0$, the horizon degenerates to a planar symmetric structure. In this paper, we choose $a = \dfrac{1}{2\pi}$ to ensure that the BH horizon has the more physically common spherical topology. This choice is consistent with the geometrical thermodynamic conclusion that $a > 0$ guarantees a spherically symmetric horizon, and also ensures consistency with the Schwarzschild BH metric. As for the molecular volume parameter $b$, its physical meaning is the same as the molecular volume of the BH in the Van der Waals fluid, and it corresponds to the macroscopic volume characterization of the microscopic BH molecules~\cite{ref5}; it likewise bears a direct connection to the characteristics of the matter field exterior to the horizon~\cite{ref52}.Combining with BH thermodynamic studies~\cite{ref53}, the BH specific volume and the horizon radius satisfy $v = 2r_+ + 3b$, which further confirms that the molecular volume parameter $b$ is an effective macroscopic parameter equivalent to the microscopic specific volume. The magnitude of the latter serves as a direct indicator of the spatial number density of the BH microscopic molecules—as $b$ increases, the microscopic volume occupied by the BH molecules expands, which at the macroscopic level is reflected in an enlargement of the horizon radius $r_h$. From Eq.~\eqref{eq:3}, it is evident that when $P = 0$ and $b = 0$, the metric function reduces to that of the Schwarzschild BH. Therefore, in this paper we will discuss the case with $b > 0$. It should also be noted that quantum gravity effects may introduce corrections to the Van der Waals BH metric, thereby affecting the fundamental properties of the photon ring~\cite{ref39,ref54}. While the present investigation is confined to the effects of the molecular volume parameter $b$ within the classical spacetime framework, the possible contributions arising from quantum corrections could point toward avenues for subsequent generalizations. On this basis, we obtain the metric equation in the following form
\begin{align}
f(r) &= 1 - \frac{2}{r} + \frac{8 \times 10^{-6}\pi}{3}r^2 \left(1 + \frac{3b}{2r}\right) \notag \\
&\quad - \frac{2b}{r} \log\left(\frac{r}{b}\right) - \frac{15b^2}{4r^2}, 
\label{eq:4}
\end{align}

That is, $f(r)$ is a function of $r$ only. Under this metric, the possible horizons are found by solving
\begin{equation}
f(r) = 0,
\label{eq:5}
\end{equation}
and the solution is denoted as $r_h = r_h(b)$, indicating the horizon radius as a function of $b$.

\begin{figure}[htbp]
    \centering
    \includegraphics[width=0.5\textwidth]{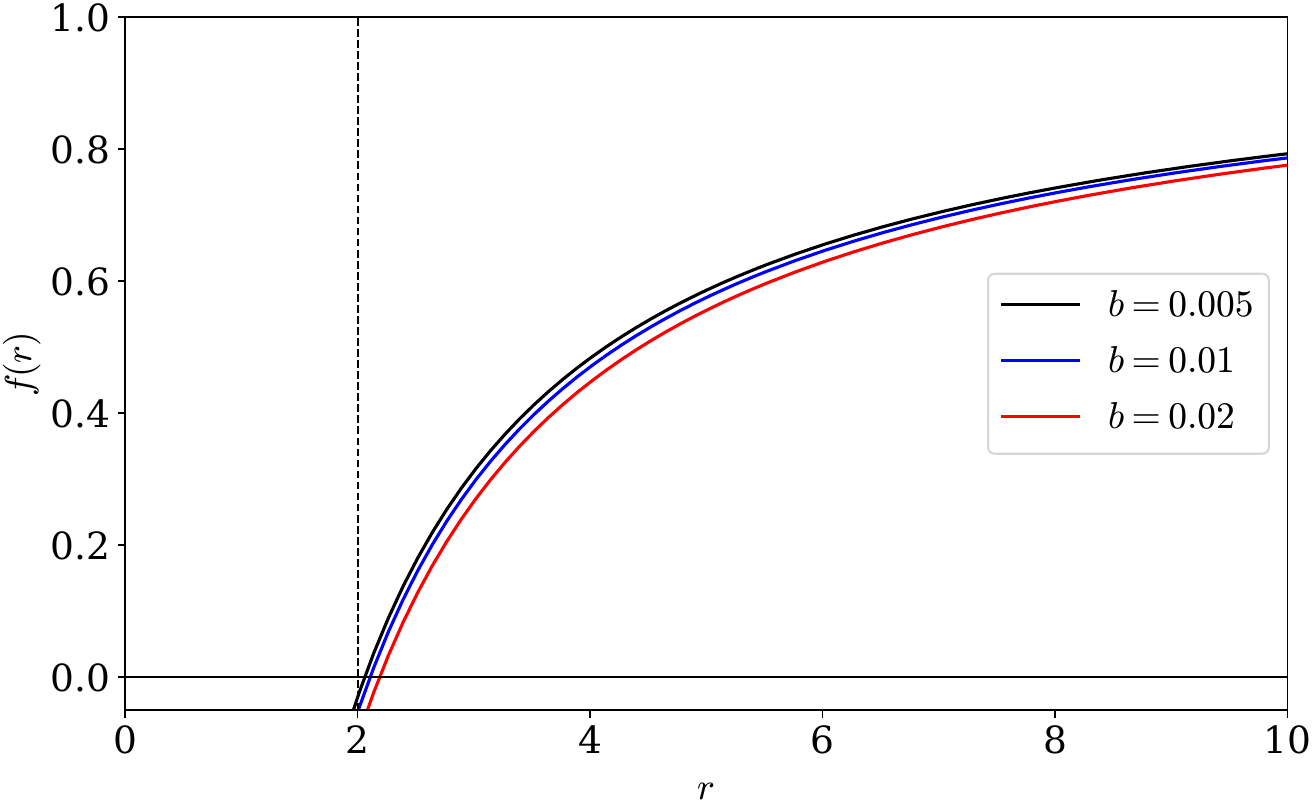}
    \caption{The metric function $f(r)$ versus $r$ for different values of $b$: $b = 0.005$ (black), $0.01$ (blue), and $0.02$ (red). T he horizon radius $r_h$ increases with increasing $b$.}
    \label{fig:1}
\end{figure}

To remain close to the Schwarzschild BH, we choose small values of $b$, specifically $b = 0.005, 0.01, 0.02$, to examine their effects on the metric function, as shown in Fig.~\ref{fig:1}.

As illustrated in Fig.~\ref{fig:1}, the magnitude of the event horizon (EH) radius—identified as the point where the curve intersects the horizontal axis—is governed by the value of $b$ : a larger $b$ leads to a correspondingly larger EH radius. When $b$ approaches zero, the Schwarzschild solution is recovered, corresponding to $r = 2$ as indicated by the black dashed line in Fig.~\ref{fig:1}. In the vicinity of the EH radius $r_h$, we can see that larger values of $b$ lead to a flatter curve, which precisely indicates that the BH molecular volume effect (short-range interaction) modifies the metric more significantly near the horizon. Furthermore, when the BH radius is very large, the curves corresponding to different values of $b$ gradually converge, implying that the molecular volume effect becomes negligible in the asymptotically AdS region, and the metric returns to the form without the molecular volume correction, as briefly discussed earlier.

Combining with thermodynamic studies~\cite{ref5}, one can further establish a qualitative connection between the BH molecular volume parameter $b$ and the microscopic number density $n$ of BH molecules through the ``BH molecule'' model.

Under this unit system, the Planck length $l_p = \sqrt{\hbar G / c^3}$ is also numerically equal to 1 (since $\hbar = 1$), so all physical quantities are dimensionless. By analogy with the microstructure of the Van der Waals fluid, Wei et al. confirmed through thermodynamic phase transitions and geometric analysis that AdS BHs possess microscopic degrees of freedom analogous to ``molecules.'' A linear relationship is obeyed between the dimensionless specific volume $v$ and the BH EH radius $r_h$
\begin{equation}
v = 2r_h + 3b. 
\label{eq:6}
\end{equation}
Defining the number of microscopic particles per unit thermodynamic volume as the dimensionless number density $n$ of BHs, its physical meaning is the spatial distribution density of BH molecules—a larger value indicates a denser packing of molecules per unit volume. It is inversely related to the specific volume $v$
\begin{equation}
n = \frac{1}{v} = \frac{1}{2r_h + 3b}.
\label{eq:7}
\end{equation}

As can be seen from the curves of the Van der Waals BH metric function, the molecular volume parameter $b$, serving as the macroscopic equivalent of the BH molecular volume, directly influences the horizon radius $r_h$. When $b$ increases, the spatial volume occupied by the BH molecules expands, which macroscopically manifests as an enlargement of the BH horizon ($r_h$ increases)~\cite{ref5}. Combined with Eq.~\eqref{eq:7}, $r_h$ is inversely related to $n$, so the qualitative relation between $b$ and the number density $n$ is: $b$ increases $\rightarrow$ $r_h$ increases $\rightarrow$ $v$ increases $\rightarrow$ $n$ decreases. This connection reveals the microscopic physical origin of the molecular volume parameter $b$---the variation of $b$ is essentially the macroscopic manifestation of the microscopic molecular volume of the BH. Through its influence on the weak attractive interactions among the BH molecules, $b$ gives rise to modifications in the spacetime curvature distribution, which in turn exerts an indirect effect on the photon trajectories within the BH gravitational field. This provides a microscopic thermodynamic foundation for the subsequent analysis of photon motion and the regulation of photon ring characteristics.

\section{THEORETICAL ANALYSIS AND OBSERVATIONAL CONSTRAINTS OF PHOTON RINGS}
\label{sec:level3}
The photon ring is one of the unique optical signatures surrounding a BH. Photons are not deflected by forces in the traditional sense, but rather propagate along null geodesics in a curved spacetime, with their trajectories appearing curved in spatial projection~\cite{ref27,ref28,ref29,ref30,ref31,ref32}. Such physical quantities of the BH photon ring as its radius and brightness distribution bear a tight connection to the corresponding spacetime metric of the BH, which provides a very effective means for probing the microscopic structure and macroscopic spacetime geometry of BHs, as well as for investigating the relevant physical properties of BHs themselves~\cite{ref13,ref27,ref42,ref55}. By combining the thermodynamic properties and geometric structure of the Van der Waals BH, this section addresses the physical mechanism underlying the photon rings, the laws of photon null geodesic motion, and the specific computational methods for the observed intensity, thereby laying the theoretical foundation for subsequent parameter constraints and the simulation of optical appearance using toy models.
\subsection{\label{sec:leve3.1}Null geodesic motion of photons}
Owing to their vanishing rest mass, photons propagate through the BH spacetime along null geodesics, whose behavior is entirely dictated by the underlying metric. Beginning with the static spherically symmetric line element of the Van der Waals BH, we employ the Lagrangian approach to obtain the equations governing photon motion, thereby extracting the associated constants of motion and constraints on the trajectories.

Photons propagate along null geodesics in spacetime. With the aim of obtaining the equations of motion, the photon Lagrangian is first formulated by employing the Van der Waals BH line element Eq.~\eqref{eq:1}
\begin{equation}
\mathcal{L} = \frac{1}{2} g_{\mu\nu} \dot{x}^\mu \dot{x}^\nu, 
\label{eq:8}
\end{equation}
where an overdot stands for the derivative taken with respect to the affine parameter $\tau$, substitution of the metric from Eq.~\eqref{eq:1} into the above expression yields
\begin{equation}
\mathcal{L} = \frac{1}{2} \left[ -f(r)\dot{t}^2 + \frac{1}{f(r)} \dot{r}^2 + r^2 \dot{\theta}^2 + r^2 \sin^2\theta \, \dot{\varphi}^2 \right]. 
\label{eq:9}
\end{equation}
From the Lagrangian, the general expression for the canonical momentum is given by
\begin{equation}
P_\mu = \frac{\partial \mathcal{L}}{\partial \dot{x}^\mu}. 
\label{eq:10}
\end{equation}
Substituting the explicit form of the Lagrangian, we obtain the individual components
\begin{align}
P_t &= \frac{\partial \mathcal{L}}{\partial \dot{t}} = -f(r) \dot{t}, \label{eq:11} \\
P_r &= \frac{\partial \mathcal{L}}{\partial \dot{r}} = \frac{1}{f(r)} \dot{r}, \label{eq:12} \\
P_\theta &= \frac{\partial \mathcal{L}}{\partial \dot{\theta}} = r^2 \dot{\theta}, \label{eq:13} \\
P_\varphi &= \frac{\partial \mathcal{L}}{\partial \dot{\varphi}} = r^2 \sin^2 \theta \, \dot{\varphi}. \label{eq:14}
\end{align}
Since neither the metric nor the Lagrangian depends explicitly on the coordinates $t$ and $\varphi$, i.e., $\partial \mathcal{L}/\partial t = \partial \mathcal{L}/\partial \varphi= 0$, it follows from the Euler--Lagrange equation
\begin{equation}
\frac{d}{d\tau} \frac{\partial \mathcal{L}}{\partial \dot{x}^\mu} = \frac{\partial \mathcal{L}}{\partial x^\mu} 
\label{eq:15}
\end{equation}
that for the cyclic coordinates $t$ and $\phi$, the corresponding canonical momenta are conserved
\begin{align}
p_t = -f(r) \dot{t} \equiv -E \quad &\Longrightarrow \quad \dot{t} = \frac{E}{f(r)},
\label{eq:16} \\
p_\varphi = r^2 \sin^2 \theta \, \dot{\varphi} = L \quad &\Longrightarrow \quad \dot{\varphi} = \frac{L}{r^2 \sin^2 \theta}.
\label{eq:17}
\end{align}
where $E$ represents the photon energy, $L$ represents the angular momentum per unit mass, and both are conserved quantities. Thanks to the spherical symmetry of the spacetime, the photon motion is restricted to the equatorial plane $\theta = \pi/2$, which gives $\sin\theta = 1$ and $\dot{\theta} = 0$. Thus we have
\begin{equation}
\dot{\varphi} = \frac{L}{r^2}. 
\label{eq:18}
\end{equation}
Photons move along null geodesics, satisfying $\mathcal{L} = 0$. Substituting Eqs.~\eqref{eq:16} and \eqref{eq:18} into Eq.~\eqref{eq:9} and restricting to the equatorial plane yields
\begin{equation}
-f(r)\left(\frac{E}{f(r)}\right)^2 + \frac{1}{f(r)} \dot{r}^2 + r^2\left(\frac{L}{r^2}\right)^2 = 0, \label{eq:19}
\end{equation}
After rearrangement, one thus obtains the equation governing the radial motion
\begin{equation}
\dot{r}^2 = E^2 - f(r) \frac{L^2}{r^2}. 
\label{eq:20}
\end{equation}
Defining the effective potential
\begin{equation}
V_{\text{eff}}(r) = \frac{f(r)}{r^2}, 
\label{eq:21}
\end{equation}
Eq.~\eqref{eq:20} can be compactly rewritten as
\begin{equation}
\dot{r}^2 = E^2 - L^2 V_{\text{eff}}(r).
\label{eq:22}
\end{equation}
To relate the radial motion to the angular motion, we define the impact parameter
\begin{equation}
k = \frac{L}{E}. 
\label{eq:23}
\end{equation}
Geometrically, this quantity represents the perpendicular distance separating the asymptotic incident direction of the photon from the BH center. Using $\dot{\varphi} = L/r^2$, we change the variable from $t$ to $\varphi$
\begin{equation}
\dot{r} = \frac{dr}{d\varphi} \dot{\varphi} = \frac{dr}{d\varphi} \frac{L}{r^2}.
\label{eq:24}
\end{equation}
Substituting this into Eq.~\eqref{eq:20}, we obtain
\begin{equation}
\left( \frac{L}{r^2} \frac{dr}{d\varphi} \right)^2 = E^2 - \frac{L^2 f(r)}{r^2}, 
\label{eq:25}
\end{equation}
Dividing both sides by $L^2/r^4$, we get
\begin{equation}
\left( \frac{dr}{d\varphi} \right)^2 = \frac{r^4}{k^2} - r^2 f(r). 
\label{eq:26}
\end{equation}
Further introducing $u = 1/r$, we have $du = -dr/r^2$, with $r^2 = 1/u^2$ and $r^4 = 1/u^4$. Substituting into Eq.~\eqref{eq:26} and simplifying, we finally obtain the standard form of the null geodesic equation
\begin{equation}
\left( \frac{du}{d\varphi} \right)^2 = \frac{1}{k^2} - u^2 f\left( \frac{1}{u} \right).
\label{eq:27}
\end{equation}
This equation fully specifies the orbital geometry of photons in the equatorial plane and provides the basis for the ensuing discussion of the effective potential, the photon ring radii, and the classification of photon rings.

\subsection{\label{sec:leve3.2}Photon Sphere and Effective Potential Analysis}
The emergence of the photon ring bears a tight connection to the photon sphere---the latter being the collection of unstable circular photon orbits within the intense gravitational field of a BH. As the primary factor determining the photon ring radius, $r_{\text{ph}}$ can be derived from the extremum condition imposed on the effective potential.

\begin{figure}[htbp]
    \centering
    \includegraphics[width=0.5\textwidth]{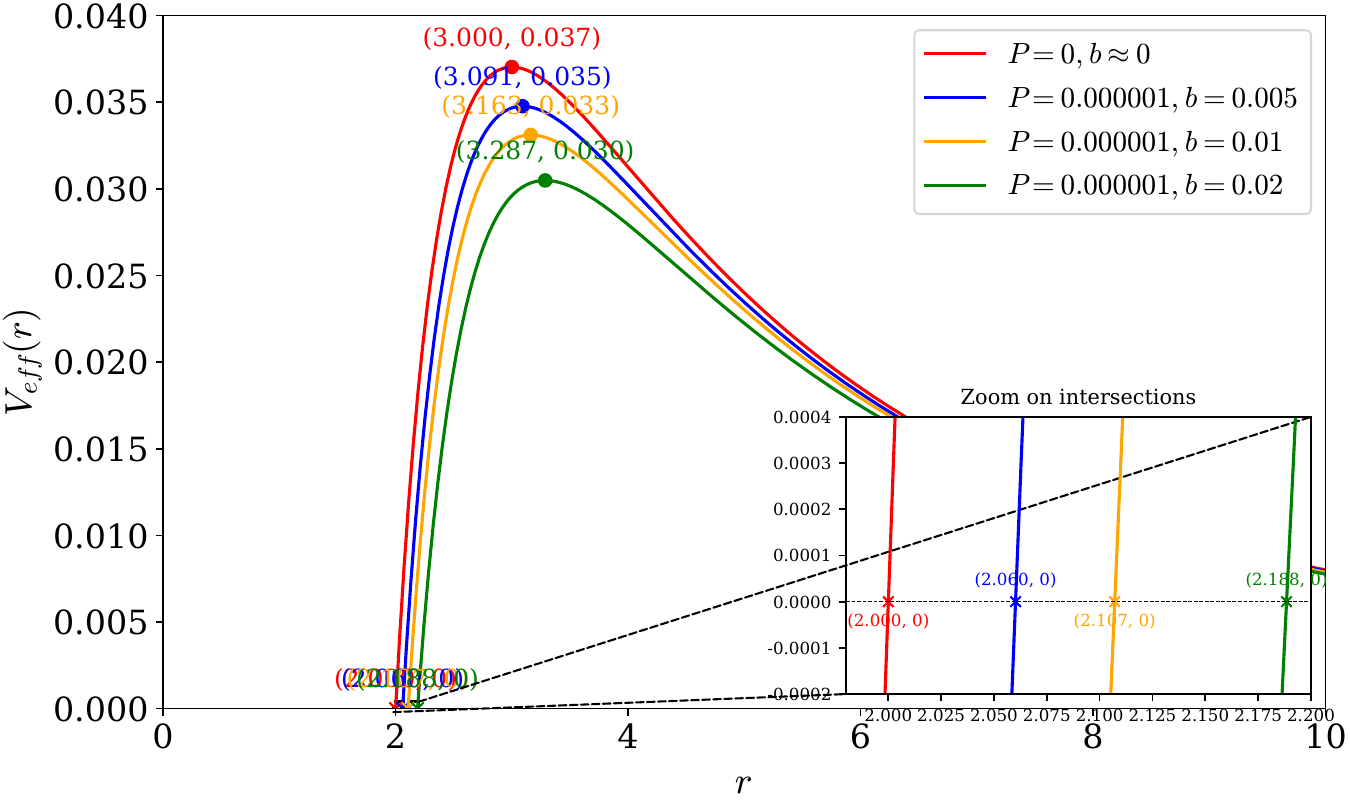}
    \caption{The effective potential $V_{\text{eff}}(r)$ as a function of $r$ for $b = 0.005, 0.01, 0.02$, and the Schwarzschild case($P=0,b \approx 0$).}
    \label{fig:2}
\end{figure}

The extremal properties of the effective potential play a role not merely in ascertaining the existence of the photon sphere, but also in reflecting, at a deeper level, the weak attractive interactions among microscopic BH molecules~\cite{ref5} and the associated thermodynamic stability~\cite{ref35}. More precisely, the change in the effective potential of the Van der Waals BH bears a direct connection to the second-order phase transition:
\begin{itemize}
    \item As the BH resides in the stable phase (the large BH phase), both the peak and the location of the effective potential tend to be more stable, which in turn makes the variation range of the photon sphere radius better aligned with observations;
    \item In the unstable phase (small BH phase), on the other hand, the effective potential exhibits larger fluctuations, and the photon ring structure is prone to deviations.
\end{itemize}
This connection can be verified by thermodynamic geometry~\cite{ref5}. The Ruppeiner scalar curvature changes sign at the critical point: $R > 0$for small BHs, indicating repulsive interactions, and $R < 0$ for large BHs, indicating attractive interactions. The repulsive nature is more pronounced for smaller BHs, whereas larger BHs progressively approach the ideal gas limit, in which the attractive interactions become weak and the interactions among BH molecules become negligible. These effective interactions, through modifications of the metric parameters, indirectly affect the spacetime structure, causing the peak position and intensity of the effective potential $V_{\text{eff}}(r)$ for photon motion to vary with the BH molecular volume parameter $b$ (as shown in Fig.~\ref{fig:2}): as the BH molecular volume parameter $b$ increases, the microscopic attractive interaction weakens, the maximum of the effective potential is lowered and displaced toward larger radii, which ultimately gives rise to an enlargement of the photon sphere radius $r_{\text{ph}}$.

It should be noted that for conventional scalar hairy BHs, the effective potential is also influenced to some extent by the scalar field potential~\cite{ref56}. Photon trajectories are fundamentally determined by the null geodesic structure of spacetime, and effects such as gravitational lensing deflection and time delay during propagation can all be regarded as manifestations of this geodesic structure in different observables~\cite{ref57,ref58}. This further confirms the crucial role of effective potential analysis in the formation of photon rings. Specifically:
\begin{enumerate}
    \item For $1/k^2 > V_{\text{eff}}(r)_{\text{max}}$, photons are either captured by the BH directly or manage to escape toward infinity, which is identified as ``direct emission'' radiation.
    \item For $1/k^2 = V_{\text{eff}}(r)_{\text{max}}$, photons follow circular trajectories at the maximum of the effective potential, giving rise to the photon ring.
    \item For $V_{\text{eff}}(r)_{\text{min}} < 1/k^2 < V_{\text{eff}}(r)_{\text{max}}$, photons orbit the BH multiple times before they finally escape, which is classified as ``photon ring'' radiation.
\end{enumerate}

Circular photon trajectories define the photon sphere. Along such trajectories, neither the radial velocity nor the radial acceleration is nonzero, i.e., $\dot{r} = 0$ and $\ddot{r} = 0$. Hence we have
\begin{equation}
\frac{1}{k^2} = V_{\text{eff}}(r). 
\label{eq:28}
\end{equation}
Differentiating Eq.~\eqref{eq:22} with respect to the affine parameter $r$, we obtain the extremum condition for the effective potential~\cite{ref59}
\begin{equation}
\left. \frac{dV_{\text{eff}}}{dr} \right|_{r = r_{\text{ph}}} = 0.
\label{eq:29}
\end{equation}
Substituting the definition of the effective potential $V_{\text{eff}}(r) = f(r)/r^2$ into Eq.~\eqref{eq:29} yields
\begin{equation}
\frac{f'(r_{\text{ph}})}{r_{\text{ph}}^2} - \frac{2f(r_{\text{ph}})}{r_{\text{ph}}^3} = 0 \quad \Longrightarrow \quad f'(r_{\text{ph}}) r_{\text{ph}} = 2f(r_{\text{ph}}).
\label{eq:30}
\end{equation}
Combining Eqs.~\eqref{eq:28} and \eqref{eq:30}, we can solve for the photon sphere radius $r_{\text{ph}}$ and the corresponding critical impact parameter $k_{\text{ph}}$~\cite{ref59}
\begin{equation}
V_{\text{eff}}(r_{\text{ph}}) = \frac{1}{k_{\text{ph}}^2},
\label{eq:31}
\end{equation}
and
\begin{equation}
V'_{\text{eff}}(r_{\text{ph}}) = 0. 
\label{eq:32}
\end{equation}
Eq.~\eqref{eq:31} serves as the fundamental starting point for the subsequent analysis of photon ring structure and shadow radius. How the effective potential $V_{\text{eff}}(r)$ varies with $r$ is displayed in Fig.~\ref{fig:2} for different values of the molecular volume parameter $b$.

The red curve in Fig.~\ref{fig:2}represents the Schwarzschild BH, for which the EH radius is $r_h = 2$ and the photon sphere radius is $r_{\text{ph}} = 3$. Fig.~\ref{fig:2} further reveals that with the growth of the BH molecular volume parameter $b$, the EH radius $r_h$ (the intercept on the $r$-axis) as well as the photon sphere radius $r_{\text{ph}}$ (the $r$-value at the peak of the effective potential) exhibit a gradual increase, whereas the peak value of the effective potential declines. Substituting this maximum value into Eq.~\eqref{eq:28} shows that a smaller maximum corresponds to a larger critical impact parameter $k_{\text{ph}}$.

\subsection{\label{sec:leve3.3}Classification of Photon Rings}
How many times a photon trajectory crosses the accretion disk governs how the emergent radiation is classified. Three distinct categories are identified: direct emission, lensing ring, and photon ring ~\cite{ref27,ref37}. Among these, the photon ring originates from photons that traverse the accretion disk at least three times, with concentrated radiation intensity and fine structure, making it the core observational target for testing BH spacetimes.

Null geodesics are classified by how often they cross the accretion disk, and their radiative contributions are grouped into three distinct categories. This classification scheme has been extensively employed in numerical simulations of BH shadows and photon rings~\cite{ref27,ref37}. With the total polar-angle variation defined as $N = \varphi / (2\pi)$, where $\varphi$ denotes the cumulative angular displacement of the photon along its trajectory from the emission point to infinity~\cite{ref27,ref54,ref58,ref60}:
\begin{enumerate}
    \item When $0 < N < 3/4$, the photon intersects the accretion disk only once and then escapes directly, referred to as direct emission, with impact parameter range $k \in (0, k_2^-) \cup (k_2^+, +\infty)$. The radiation intensity is diffuse.
    \item In the interval $3/4 < N < 5/4$, a photon crosses the accretion disk on two occasions and then escapes after being bent by the BH's gravitational lensing; this case is identified as the lensed ring, with impact parameter lying in $k \in (k_2^-, k_3^-) \cup (k_3^+, k_2^+)$ and giving rise to a secondary bright ring.
    \item When $N > 5/4$, the photon revolves around the BH for more than 1.25 turns before escaping, intersecting the accretion disk at least 3 times. The corresponding impact parameter range is $k \in (k_3^-, k_3^+)$, where $k_3^-$ and $k_3^+$ are the boundaries of the impact parameter for the photon ring. The radiation intensity is concentrated at its peak, forming a thin and bright photon ring.
\end{enumerate}

\begin{figure}[htbp]
    \centering
    \includegraphics[width=0.5\textwidth]{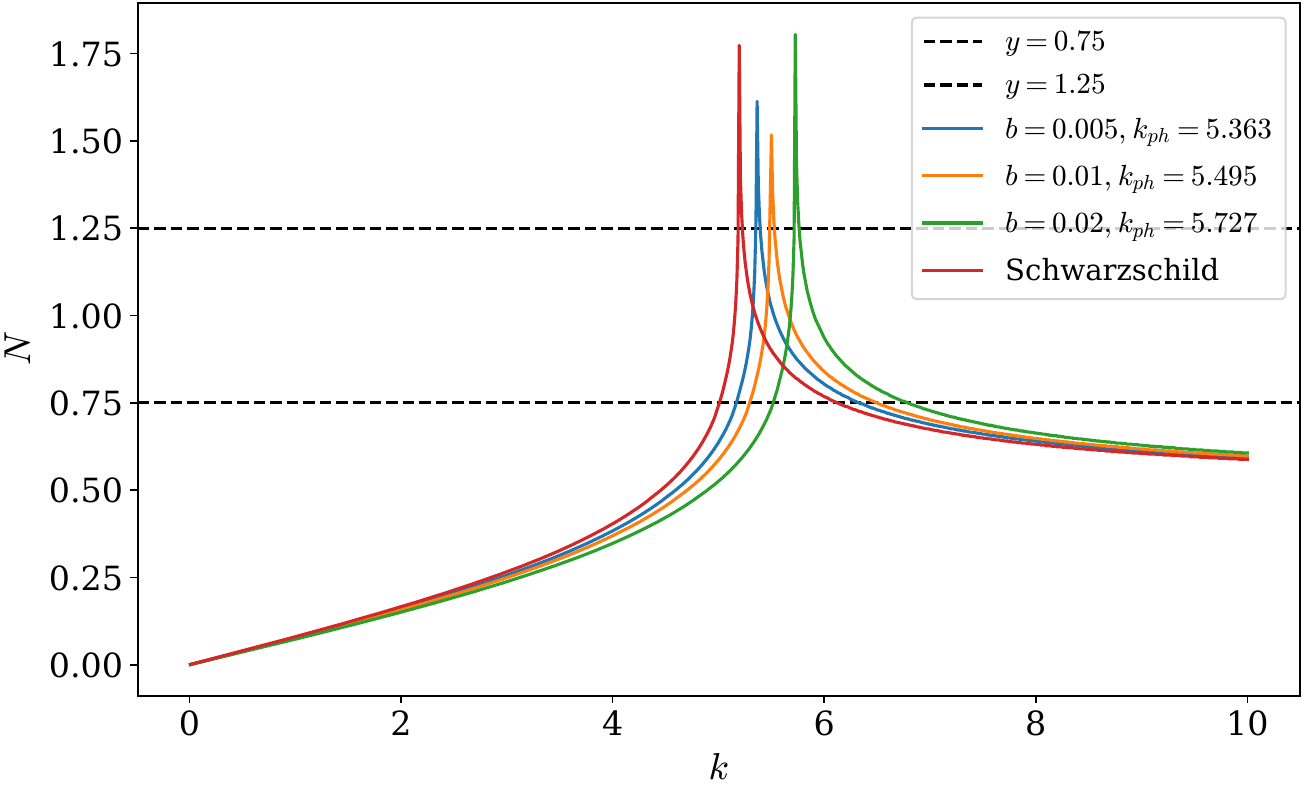}
    \caption{Dependence of $N$ on $k$. Curves are displayed for $b = 0.005, 0.01, 0.02$, as well as for the Schwarzschild case ($P=0,b \approx 0$).}
    \label{fig:3}
\end{figure}

With the help of the null geodesic Eq.~\eqref{eq:27}, the relation between $N$ and $k$ is derived, which is displayed in Fig.~\ref{fig:3}. Here, the BH molecular volume parameter $b$ is chosen as $b = 0.005, 0.01, 0.02$, together with the Schwarzschild solution, for a four-case comparative study.

From Fig.~\ref{fig:3}, it can be seen that in all cases the bending of the null geodesic diverges at the critical impact parameter $k = k_{\text{ph}}$. For $k < k_{\text{ph}}$, the bending increases monotonically with $k$ and diverges as $k \to k_{\text{ph}}^{-}$; for $k > k_{\text{ph}}$, the bending decreases monotonically from infinity as $k$ increases further. In addition, we observe that as the BH molecular volume parameter $b$ decreases, $k_{\text{ph}}$ also decreases, eventually approaching the value $3\sqrt{3}$ corresponding to the Schwarzschild BH. The essence of this behavior lies in the relation between $b$ and the BH molecular number density $n$: a decrease in $b$ corresponds to a smaller molecular volume, which implies a larger number density $n = 1/(2r_h+3b)$, so that the BH microstructure approaches the uniform distribution of the Schwarzschild BH. Consequently, the critical impact parameter $k_{\text{ph}}$ of the photon sphere also tends to the Schwarzschild theoretical value.

It is noteworthy that the impact parameter boundaries of the photon ring, specifically $k_2$ and $k_3$, bear a tight correlation with the BH molecular number density $n$~\cite{ref5}. Based on the preceding qualitative analysis, when the BH molecular volume parameter $b$ increases, the BH molecules become more sparsely distributed, and the number density $n$ correspondingly decreases. As the molecular distribution becomes sparser, the originally weak attractive interactions among them become even weaker, which reduces the degree of orbital deflection of photons propagating around the BH. When the orbital deflection is less pronounced, the critical impact parameter $k_{\text{ph}}$ associated with stable photon circulation becomes larger, a point that will be further corroborated by the calculations presented below.

Hence, we are able to carry out additional precise computations to ascertain the limits of the $k$ that correspond to various numbers of crossings with the accretion disk. Meanwhile, the innermost stable circular orbit (ISCO) radius $r_{\text{isco}}$ can be obtained via the following formula~\cite{ref61}
\begin{equation}
r_{\text{isco}} = \frac{3f(r_{\text{isco}}) \cdot f'(r_{\text{isco}})}{2\left[f'(r_{\text{isco}})\right]^2 - f(r_{\text{isco}}) \cdot f''(r_{\text{isco}})},
\label{eq:33}
\end{equation}
these findings, together with those for the $r_h$, the $r_{\text{ph}}$, and the  $k_{\text{ph}}$, are compiled in Table~\ref{tab:1}.

\begin{table*}[t]
\centering
\caption{Summary of parameters $k_2^-$, $k_2^+$, $k_3^-$, $k_3^+$, $r_h$, $r_{\text{ph}}$, $k_{\text{ph}}$, and $r_{\text{isco}}$ for different values of $b$, together with the Schwarzschild case($P=0,b \approx 0$).}
\label{tab:1}
\begin{ruledtabular}
\begin{tabular}{ccccccccc}
 $b$ &$k_2^-$ &$k_2^+$ &$k_3^-$ &
 $k_3^+$ &$r_h$ &$r_{\text{ph}}$ &$k_{\text{ph}}$  &$r_{\text{isco}}$ \\
\hline
0.005 & 5.1698 & 6.3270 & 5.3535 &5.3952 & 2.0602 & 3.0923 & 5.3629 &6.1727 \\
0.01  & 5.2962 & 6.4962 & 5.4840 &5.5288 & 2.1071 & 3.1628 & 5.4949 &6.3038 \\
0.02  & 5.5175 & 6.7893 & 5.7159 &5.7630 & 2.1884 & 3.2870 & 5.7269 &6.5245 \\
Schw. & 5.0200 & 6.1700 & 5.1900 &5.2300 & 2.0000 & 3.0000 & 5.1962 &6.0000 \\
\end{tabular}
\end{ruledtabular}
\end{table*}

Combining the numerical values of the parameters in Table~\ref{tab:1} and substituting them into Eq.~\eqref{eq:7}, we obtain the corresponding values of the number density $n$, as shown in Table ~\ref{tab:2}.

\begin{table}[t]
\centering
\caption{The specific volume $v$ and number density $n$ for different values of $b$, together with the Schwarzschild case($P=0,b \approx 0$).}
\label{tab:2}
\begin{ruledtabular}
\begin{tabular}{cccc}
 $b$ & $r_h$ & $v$ & $n$ \\
\hline
0.005 & 2.0602 & 4.1354 & 0.2418 \\
0.01 & 2.1071 & 4.2442 & 0.2356 \\
0.02 & 2.1884 & 4.4368 & 0.2254 \\
Schw. & 2.0000 & 4.0000 & 0.2500 \\
\end{tabular}
\end{ruledtabular}
\end{table}

This quantitative result is consistent with the preceding qualitative analysis: as $b$ increases from 0.005 to 0.02, the number density $n$ decreases from 0.2418 to 0.2254, exhibiting a clear negative correlation. Thus, one can readily observe that when the molecular volume parameter $b$, which characterizes the microscopic molecular volume of the BH, gradually increases, the spatial distribution of these microscopic BH molecules becomes more dilute, and the corresponding number density $n$ decreases accordingly. Consequently, the originally weak attractive interactions among the BH molecules are further weakened~\cite{ref5,ref44,ref53}. Such a change in the intermolecular interactions directly affects how spacetime is structured near the BH horizon, thereby making the spacetime curvature gradient less steep ~\cite{ref61}. This in turn alters the effective potential that photons encounter as they travel around the BH---the effective potential reaches a lower peak that is shifted outward, as is evident in Fig.~\ref{fig:2}. Given that the photon sphere radius is exactly dictated by the position of the effective potential peak, the outward displacement of the peak naturally gives rise to an enlargement of the photon sphere radius $r_{\text{ph}}$, in full agreement with the numerical results listed in Table~\ref{tab:1}.

\begin{figure*}[!htbp]
    \centering
    \begin{minipage}[b]{0.45\textwidth}
        \centering
        \includegraphics[width=\textwidth]{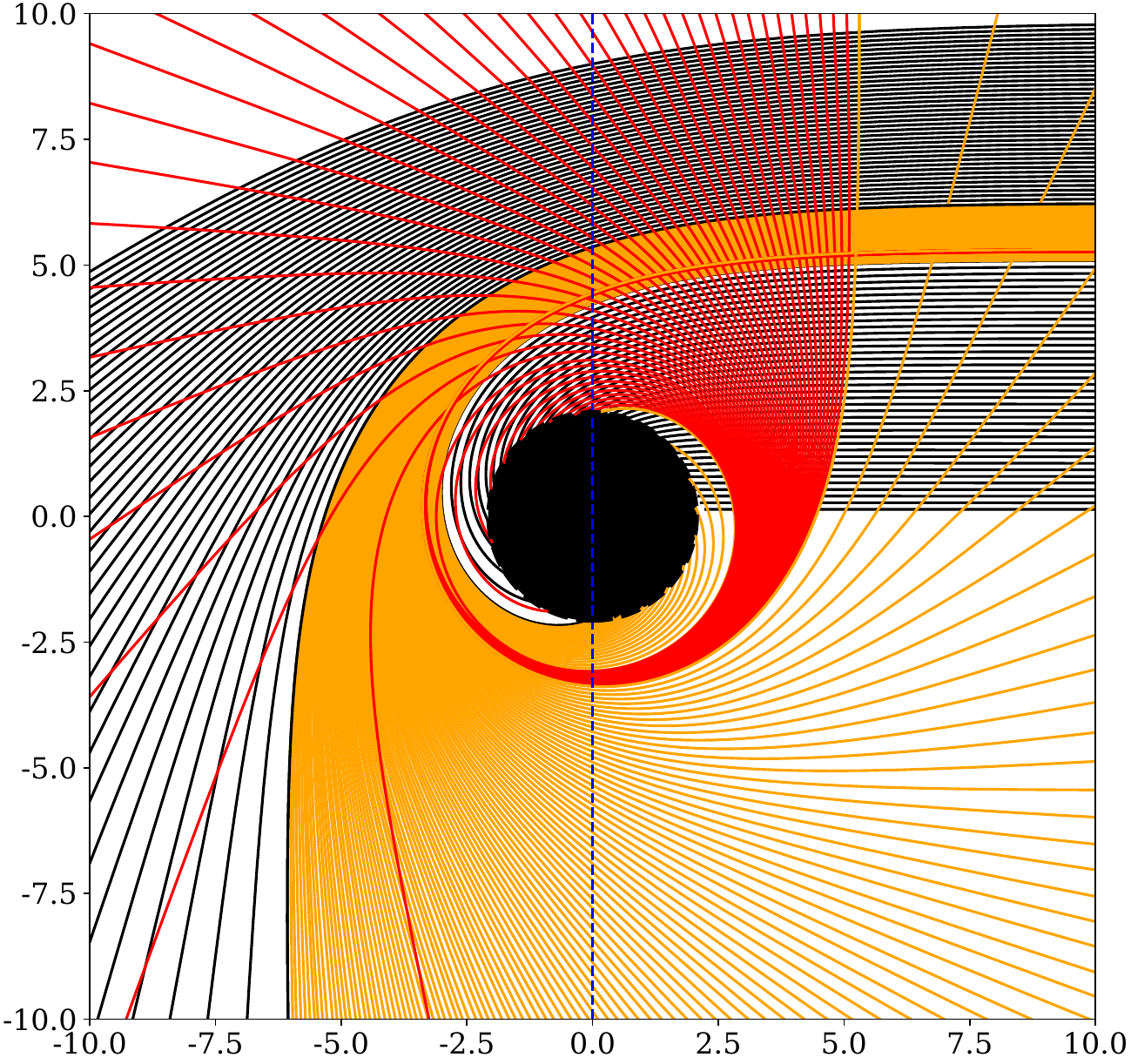}
        \textbf{(a) $0.005$}   
    \end{minipage}
    \hfill
    \begin{minipage}[b]{0.45\textwidth}
        \centering
        \includegraphics[width=\textwidth]{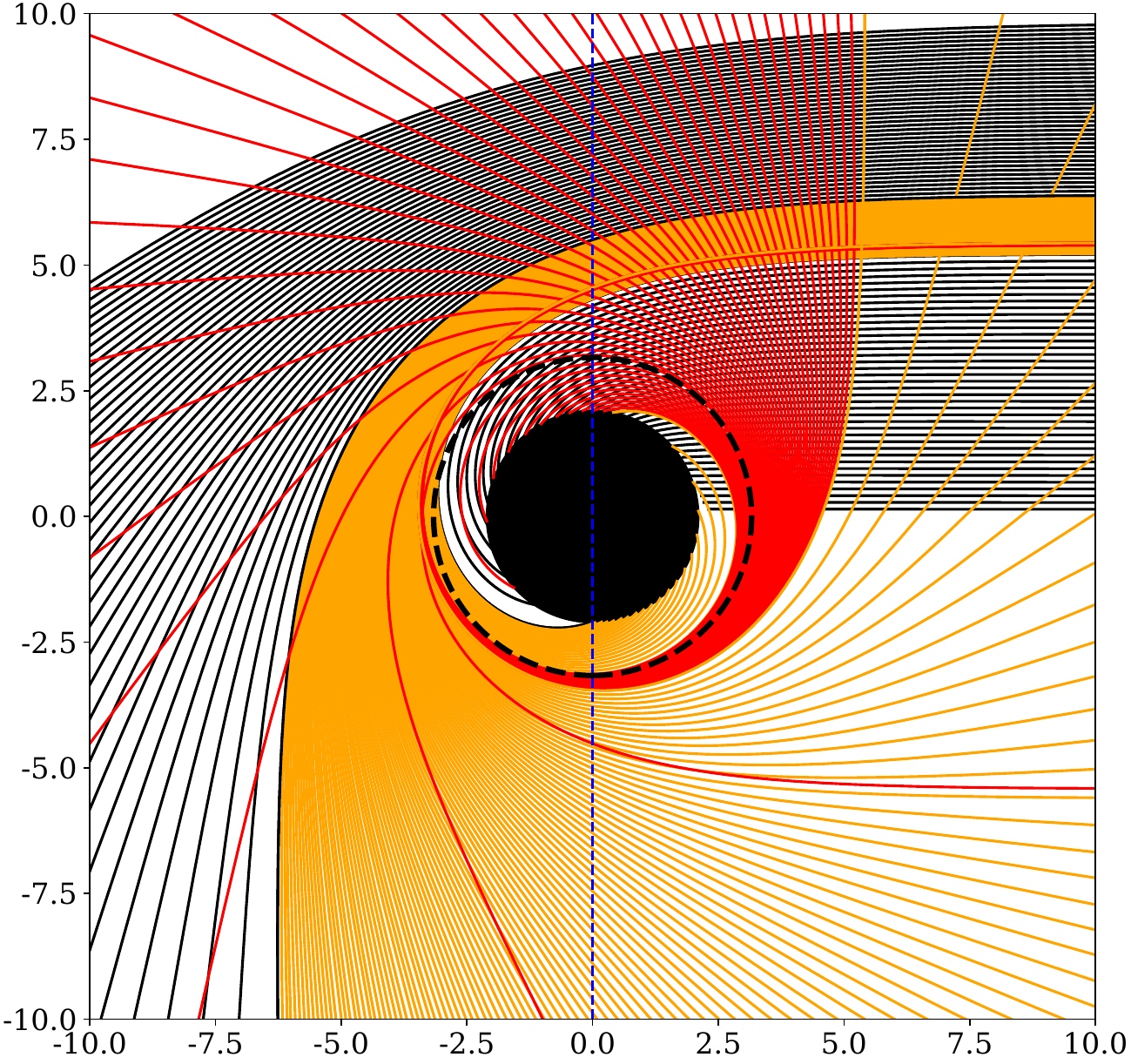}
        \textbf{(b) $0.01$}
    \end{minipage}
    \vspace{0.2cm}   
    \begin{minipage}[b]{0.45\textwidth}
        \centering
        \includegraphics[width=\textwidth]{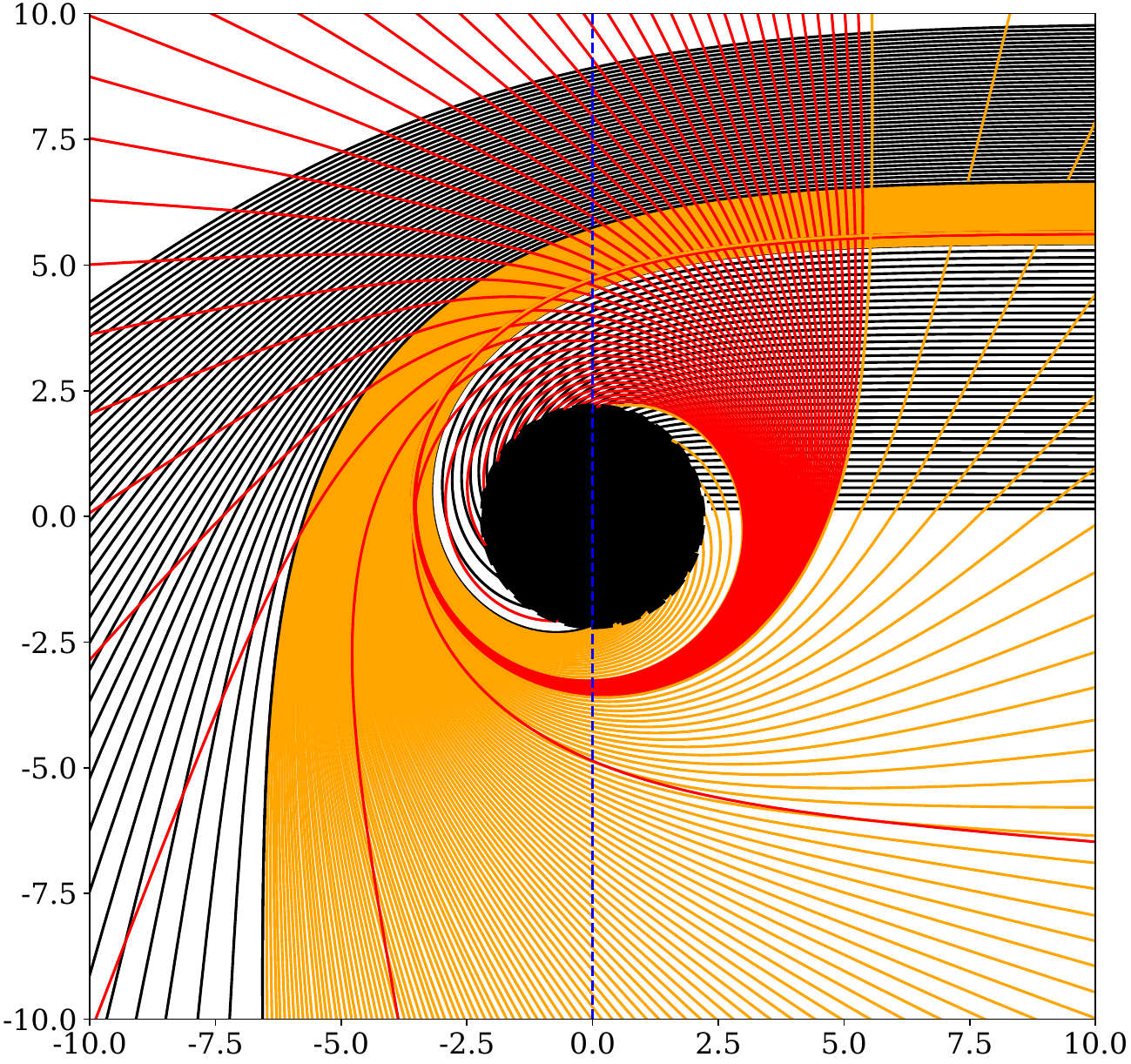}
        \textbf{(c) $0.02$}
    \end{minipage}
    \hfill
    \begin{minipage}[b]{0.45\textwidth}
        \centering
        \includegraphics[width=\textwidth]{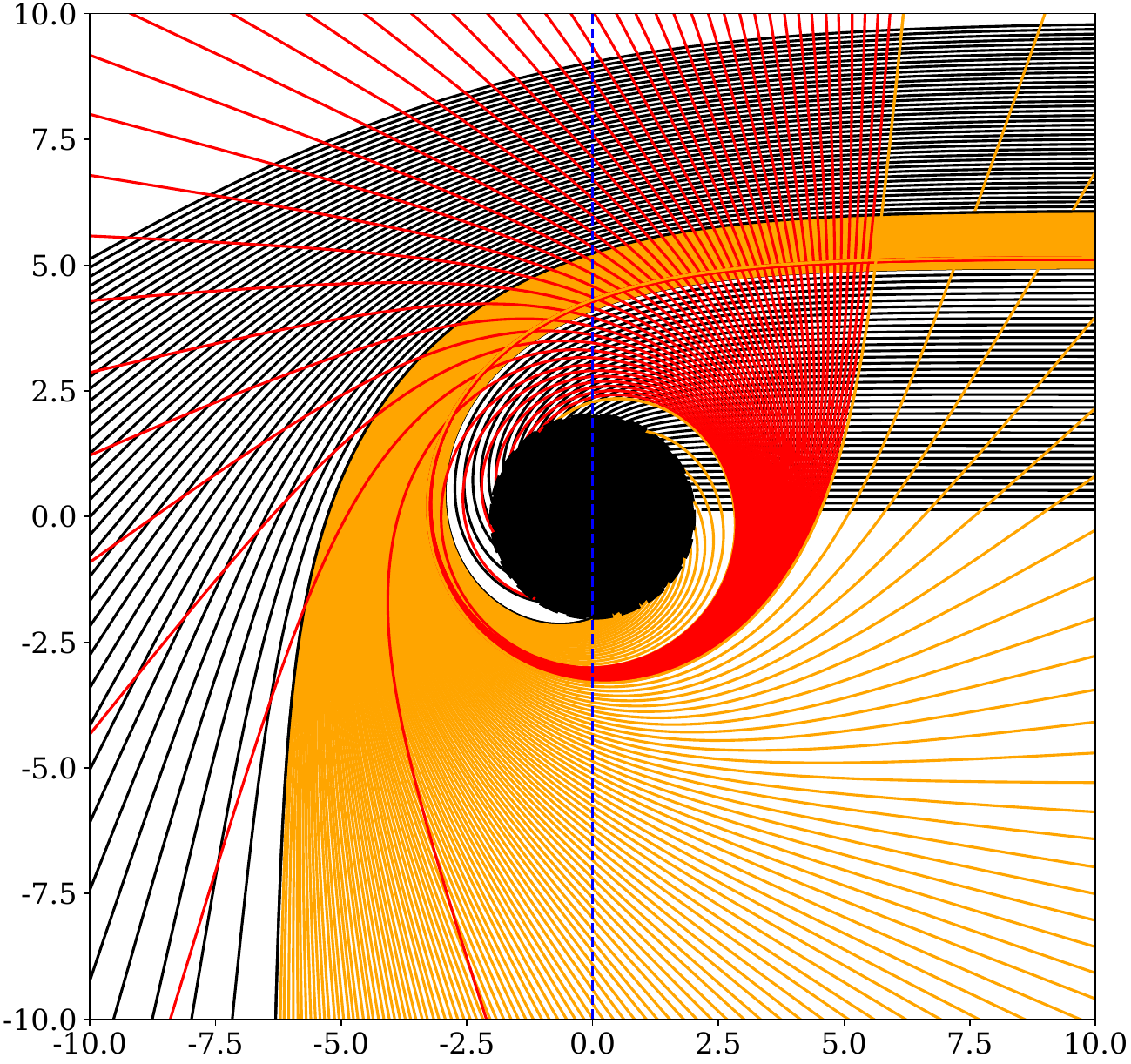}
        \textbf{(d) Schwarzschild}
    \end{minipage}
    \caption{Null geodesic trajectories for different values of $b$: (a) $0.005$, (b) $0.01$, (c) $0.02$, (d) Schwarzschild ($P=0,b \approx 0$).}
    \label{fig:4}
\end{figure*}

On the basis of the numerical results, the null geodesic equation is employed, and under the relevant physical assumptions, the null geodesic trajectories are plotted, as presented in Fig.~\ref{fig:4}.

In Fig.~\ref{fig:4}, panels (a)--(d) display the null geodesic trajectories for four cases: the BH molecular volume parameter $b = 0.005$(a), $0.01$(b), $0.02$(c), and the Schwarzschild solution (d), which is restored once the pressure parameter $P$ and the molecular volume parameter $b$ both approach zero. From Fig.~\ref{fig:4}, we observe that as the molecular volume parameter $b$ gradually increases, the corresponding Van der Waals BH $r_h$,  $r_{\text{ph}}$, $k_{\text{ph}}$ all increase monotonically, which corroborates the numerical results tabulated in Table~\ref{tab:1}. Notably, clear differences can be seen when comparing panels (a) with (c), and (c) with (d). Furthermore, by examining the trajectory variations, we find that the molecular volume parameter $b$ modifies the curvature of the spacetime around the BH, thereby affecting the photon trajectories.

\subsection{\label{sec:leve3.4}Parameter Constraints Based on EHT Observations}
The EHT measurements of the photon ring structures associated with M87* and Sgr A*~\cite{ref18,ref19,ref41,ref43} impose constraints on the parameters of theoretical BH models. Through the employment of the essential observational data made available by the EHT, including the BH shadow diameter and the photon ring morphology, in conjunction with the relationship between the observed data and the thermodynamic-geometric parameters of the Van der Waals BH, one can determine the physical meaning and feasibility of the parameters in this theoretical model.

Measurements from the EHT collaboration have yielded the angular size and geometric scale of the BH shadow, which directly reflect the horizon scale and the spacetime geometry of the strong-field region. These observational data can be connected to the Van der Waals BH model through the following steps:

\textbf{(1) Theoretical expression for the shadow radius}

The BH shadow denotes the dark area formed by photons captured in the intense gravitational field, with its radius determined jointly by the photon sphere radius and the spacetime metric. For the Van der Waals BH, by combining the features of the static spherically symmetric metric with the investigation of photon trajectories, the shadow radius $r_{\text{sh}}$ observed by a faraway static observer obeys the following relation ~\cite{ref27,ref62}
\begin{equation}
r_{\text{sh}} = \frac{r_{\text{ph}}}{\sqrt{f(r_{\text{ph}})}}.
\end{equation}
Due to the spherical symmetry of the BH shadow, the shadow diameter $d_{\text{sh}}$ is related to the shadow radius by $d_{\text{sh}} = 2r_{\text{sh}}$.

\textbf{(2) Standardization of observational data}

The key observational data released by the EHT collaboration~\cite{ref18,ref19}, after standardization, are as follows:

\begin{enumerate}
    \item \textbf{M87*}: For M87*, the EHT collaboration reported a shadow angular diameter of $42 \, \mu\text{as}$ (with $3 \, \mu\text{as}$ uncertainty), an Earth-based distance of $16.8 \, \text{Mpc}$ (with $+0.8/-0.7 \, \text{Mpc}$ uncertainty), and a mass of $6.5 \times 10^9 \, M_\odot$ (with an error of $0.9 \times 10^9 \, M_\odot$). Adopting these values, the physical diameter is $d_{\text{sh}}^{M87^*}=(11 \pm 1.5)$.

    \item \textbf{Sgr A*}: For Sgr A*, the EHT collaboration reported a shadow angular diameter of $48.7 \, \mu\text{as}$ (with $7 \, \mu\text{as}$ uncertainty), an Earth-based distance of $8277 \, \text{pc}$ (uncertain by $33 \, \text{pc}$), and a mass of $4.3 \times 10^6 \, M_\odot$ (with an error of $0.013 \times 10^6 \, M_\odot$). Adopting these values, the physical diameter is $d_{\text{sh}}^{Sgr A^*}=(9.5 \pm 1.4)$ ~\cite{ref45}.
\end{enumerate}

\begin{figure*}[!htbp]
    \centering
    \begin{minipage}[b]{0.45\textwidth}
        \centering
        \includegraphics[width=\textwidth]{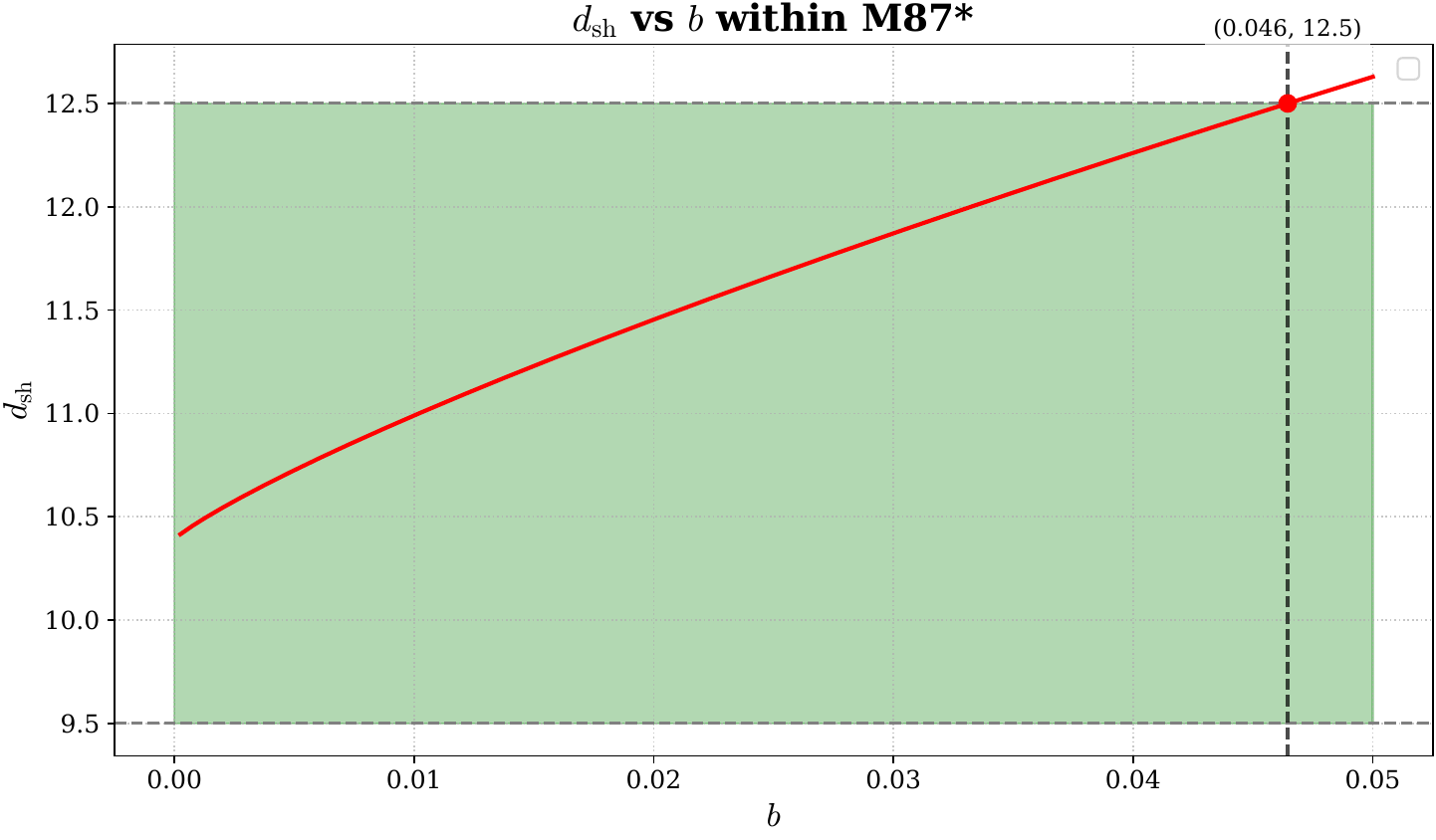}
        \textbf{(a) M87*}
    \end{minipage}
    \hfill
    \begin{minipage}[b]{0.45\textwidth}
        \centering
        \includegraphics[width=\textwidth]{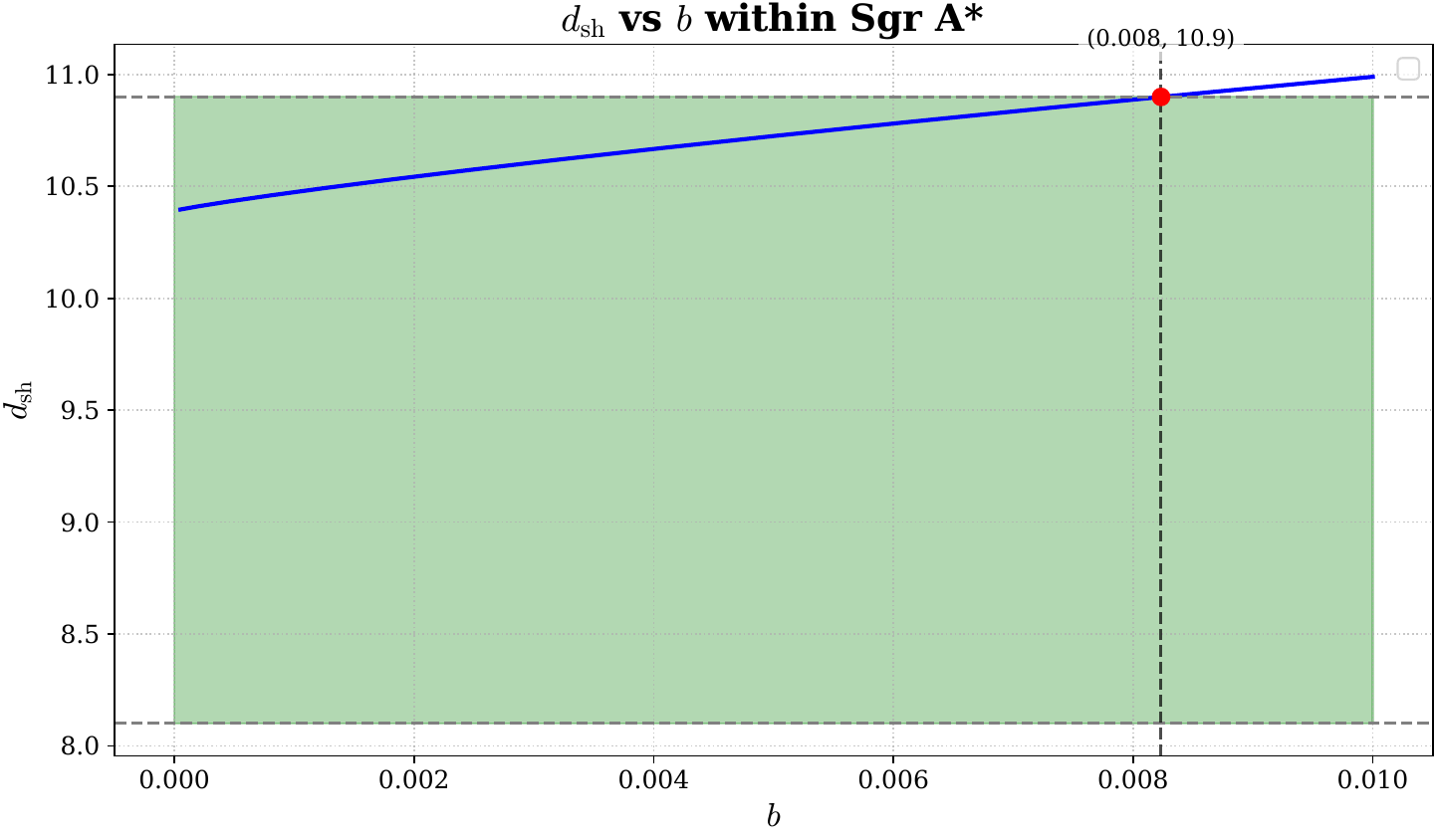}
        \textbf{(b) Sgr A*}
    \end{minipage}
    \caption{Constraints imposed on the molecular volume parameter $b$ by the EHT measurements of the shadow diameters of M87* and Sgr A*. The variation of the BH shadow diameter with $b$ is depicted by the red and blue curves, while the green shaded band denotes the observationally allowed region, from which the viable values of $b$ are inferred.}
    \label{fig:5}
\end{figure*}

Through the union of the measured shadow diameters with the theoretical formulations, along with the adoption of the parameter-constraining procedures developed for quantum-corrected BHs and regular BHs carrying magnetic charge ~\cite{ref58,ref63}, and for regular BHs in asymptotically safe gravity ~\cite{ref46}, while also incorporating the parameter bounds dictated by the energy conditions for the Van der Waals BH ~\cite{ref52} and the demand for a stable phase transition region ~\cite{ref5}, the constraints on the BH molecular volume parameter $b$ are derived, as illustrated in Fig.~\ref{fig:5}. Combined with the latest EHT photon ring morphology data~\cite{ref41,ref43,ref68}, the allowed range of $b$ can be further narrowed. This constrained range simultaneously satisfies the geometric matching with observational data, thermodynamic stability, and physical realizability requirements.

As shown in Fig.~\ref{fig:5}, the EHT-measured shadow diameters of M87* and Sgr A* impose stringent observational constraints on the BH molecular volume parameter $b$ in the Van der Waals BH model considered in this paper. The analysis reveals that, at the confidence level corresponding to the M87* shadow diameter $d_{\text{sh}}^{M87^*}=(11 \pm 1.5)$, the molecular volume parameter $b$ is constrained to the range $0 \leq b \leq 0.046$; while at the confidence level corresponding to the Sgr A* shadow diameter $d_{\text{sh}}^{Sgr A^*}=(9.5 \pm 1.4)$, the molecular volume parameter $b$ is constrained to the range $0 \leq b \leq 0.008$. Clearly, the upper bound on $b$ set by Sgr A* is more stringent than that by M87*, whereas the error range provided by M87* is wider.

It is noteworthy that the M87* constraint encompasses the range $b \in [0, 0.02]$ primarily investigated in this paper, whereas the Sgr A* constraint only encompasses $b \in [0, 0.008]$, and when $P=0$, and $b \approx 0$, the model naturally reduces to the Schwarzschild BH. Furthermore, combined with the energy condition constraints~\cite{ref52}, the value of the molecular volume parameter $b$ must also satisfy $r_h >0.898b$ to ensure $M > 0$ and the weak energy condition. The horizon radii corresponding to our chosen values $b = 0.005, 0.01, 0.02$ are 2.0602, 2.1071, and 2.1884, respectively (see Table~\ref{tab:1}), all of which satisfy $r_h > 0.898b$. Moreover, the pressure $P = 1 \times 10^{-6}$ lies within the effective range that satisfies the energy conditions~\cite{ref52}. The analysis of Ref.~\cite{ref52} reveals that the energy conditions of the Van der Waals BH depend sensitively on the pressure. For sufficiently low $P$, the stress-energy tensor satisfies all three energy conditions—the weak, the strong, and the dominant—at the same time. With rising $P$, the dominant energy condition is the first to break down, followed by the weak energy condition, whereas the strong one remains valid throughout. In this paper, we adopt a small pressure value $P = 1 \times 10^{-6}$ to ensure, to the greatest extent possible, that the model satisfies all three energy conditions outside the horizon, further validating the physical rigor of our parameter selection~\cite{ref40}.

\section{OBSERVED INTENSITY AND OPTICAL APPEARANCE SIMULATION}
\label{sec:level4}
Next, for distinct values of the molecular volume parameter $b$, numerical simulations are performed to generate the two-dimensional images of the shadow and the photon ring associated with the Van der Waals BH, as viewed from a remote observational position.

Based on Liouville's theorem and the gravitational redshift effect, the monochromatic specific intensity measured by the observer satisfies~\cite{ref55,ref66,ref67}
\begin{equation}
I_0(r, v_0) = g^3 A = g^3 I_e(r, v_e).
\label{eq:35}
\end{equation}
Here, the redshift factor is defined as $g = \dfrac{v_0}{v_e} = \sqrt{f(r)}$, which characterizes the modification of photon frequency induced by the strong gravitational field, with $v_0$ and $v_e$ denoting the observed and emitted frequencies, respectively. Once the integration over the entire frequency range is carried out, the observed intensity in total is obtained by accumulating the contributions arising from every successive crossings between the photon trajectory and the disk of accreting matter. Consequently, the net intensity detected by the observer is given by the cumulative sum over these individual intersections~\cite{ref27}
\begin{equation}
I_{\text{obs}}(k) = \sum_m \bigl[ f(r_m(k)) \bigr]^2 I_{\text{em}}(r_m(k)).
\label{eq:36}
\end{equation}
In this context, $I_{\text{em}}(r) =\int A \, dv_e$, $A=I_e(r, v_e) $. This quantity $I_{\text{em}}(r)$ represents the overall emission intensity originating from the accretion disk, reflecting the dependence of the emitted intensity on the distance from the BH; $r_m(k)$ denotes the transfer function. For a photon with the impact parameter $k$, the radial coordinate corresponding to its $m$-th crossing of the accretion disk is provided by this function. For a fixed $m$, its slope $dr_m/dk$ characterizes the strength of the decoherence amplification effect, with larger $m$ typically indicating more significant decoherence effects~\cite{ref37,ref68}.

Indeed, as revealed by the null geodesic distribution in Fig.~\ref{fig:4}), for $m > 3$ (associated with the red curve), the $k$ exerts a marked influence on the radial coordinate at the point where the geodesic crosses for the third time ~\cite{ref37,ref68}. In other words, within the neighborhood of $k$, the geodesics corresponding to impact parameters $k_1^+$ and $k_2^+$ extract markedly different energies from the accretion disk at their third intersections (as will be shown below, the radiation intensity is directly related to this radial distance). This phenomenon is precisely the significant decoherence amplification effect. Special attention should be paid to the fact that Eq.~\eqref{eq:36} is formulated with the aim of faithfully capturing the physical appearance of the photon shadow and the photon ring region of the BH when viewed from infinity. Theoretically, in the limit of infinity, the $k$ essentially encodes the radial separation between the bright arc and the BH center, a finding that has been confirmed and extensively acknowledged in the literature ~\cite{ref37,ref68}.

According to the above definitions, three physical processes can be distinguished: when $m \leq 1$, direct emission exists; when $m = 2$, the lensed ring mode is present; and when $m \geq 3$, the photon ring mode is present. When $m > 3$, the corresponding higher-order geodesics experience significantly enhanced signal attenuation, and their contribution to the total luminosity tends to zero. Provided that the radiative properties inside the accretion disk region stay in relative equilibrium with the surrounding environment, the target brightness as measured is chiefly governed by direct emission ~\cite{ref27}. The quantitative relationships among these three radiation transfer functions can be expressed by the following formulas
\begin{align}
r_1(k) &= \frac{1}{u\left(\frac{\pi}{2}, k\right)}, \quad k \in (k_1^-, +\infty), \label{eq:r1} \\
r_2(k) &= \frac{1}{u\left(\frac{3\pi}{2}, k\right)}, \quad k \in (k_2^-, k_2^+), \label{eq:r2} \\
r_3(k) &= \frac{1}{u\left(\frac{5\pi}{2}, k\right)}, \quad k \in (k_3^-, k_3^+). \label{eq:r3}
\end{align}
where $u(\varphi, k)$ denotes the solution of the geodesic equation Eq.~\eqref{eq:27}.

\begin{figure}[!htbp]
    \centering
    \includegraphics[width=0.5\textwidth]{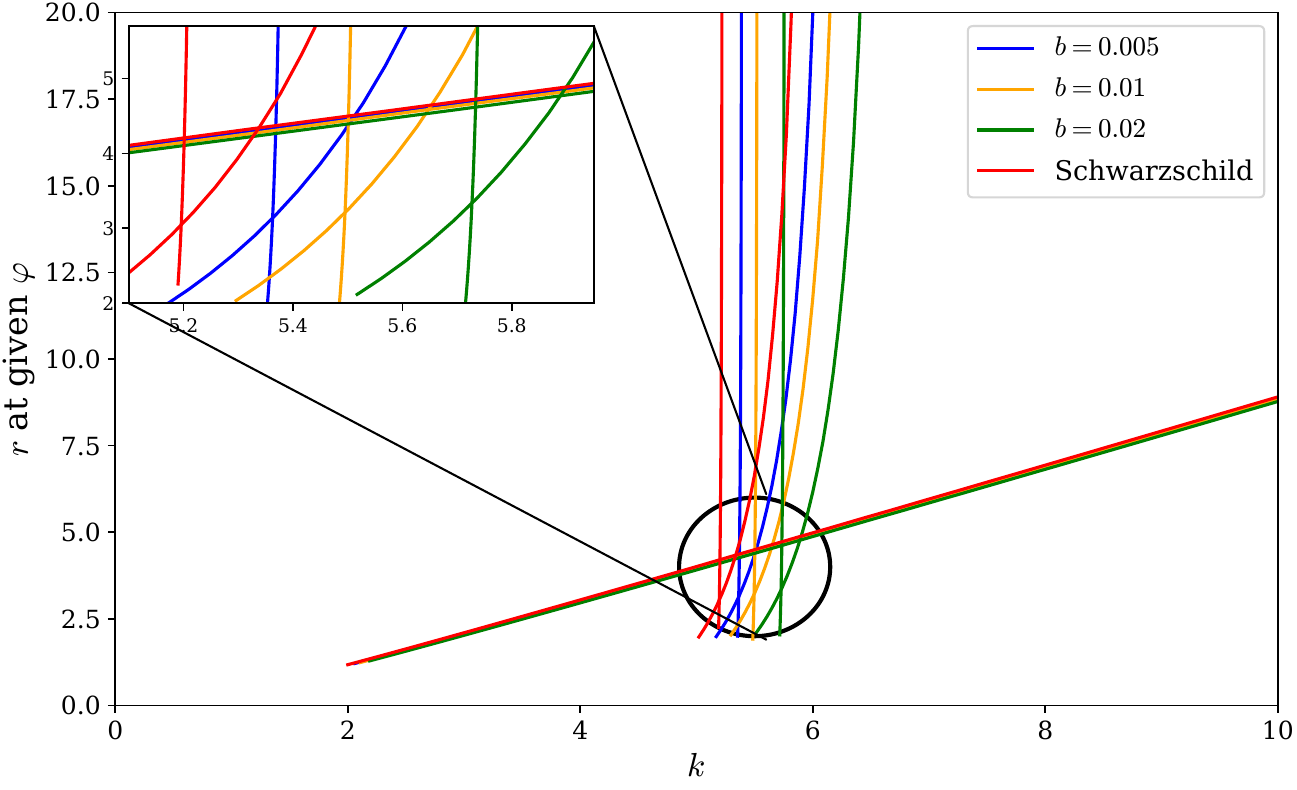}
    \caption{The first three transfer functions corresponding to $b = 0.005$ (blue), $0.01$ (orange), $0.02$ (green), and the Schwarzschild case($P=0,b \approx 0$)(red).}
    \label{fig:6}
\end{figure}

In this study, we consider three cases with the BH molecular volume parameter $b = 0.005, 0.01, 0.02$, along with the Schwarzschild solution, to analyze the spatial distribution characteristics of the transfer functions within the Schwarzschild BH background. The corresponding transfer functions are displayed in Fig.~\ref{fig:6}.

Our investigation demonstrates that with the growth of the BH molecular volume parameter $b$, the span of the transfer functions widens progressively, and the departure from the theoretical curves associated with the Schwarzschild solution grows accordingly. Across different simulation scenarios, variations in $m$ markedly steepen the transfer function slopes, thereby mirroring the impact of strong gravitational lensing on the radial coordinate. A more in-depth examination reveals that the slopes of each $r_i(k)$ approach unity. This indicates a well-behaved linear dependence between the impact parameter and the radial coordinate at the point where the null geodesic makes its first crossing of the accretion disk.

With the aim of probing the optical look of the Van der Waals BH in more depth, three canonical toy emission models are employed, which have been widely utilized in the literature concerning BH imaging ~\cite{ref27,ref58,ref62}. Earlier investigations have demonstrated that the photon ring of a rotating BH is governed by the interplay between parameters and spin ~\cite{ref38}; the optical appearance of extremal regular BHs and quantum-corrected BHs displays distinctive characteristics that depend on the accretion disk type and quantum effects ~\cite{ref60,ref69}; and analogous features have likewise been identified in the optical appearance of regular BHs within asymptotically safe gravity ~\cite{ref46}. These provide directions for subsequent comparisons of observational differences among different types of BHs.

In reality, the radiation from accretion disks does not follow a simple power-law decay. As early as 1991, Laor~\cite{ref70} proposed a broken power-law emissivity model to explain the broad emission line profiles of active galactic nuclei. In recent years, this idea has been introduced into simulations of BH photon rings. Desire et al.~\cite{ref71} constructed a physically motivated multi-frequency emission model, achieving a more accurate characterization of the coupling effects between the photon ring and the accretion disk. Numerical simulations in different gravitational frameworks have also confirmed that adopting different emission profiles significantly affects the BH shadow and photon ring~\cite{ref72}. These advances provide an important foundation for improving the realism of optical appearance simulations.

Nevertheless, with the goal of acquiring a preliminary insight into the photon ring properties of the Van der Waals BH, three toy models featuring power-law decay, which are adequate for rudimentary simulations, are employed ~\cite{ref27,ref58,ref62}. The selection criteria for the emission models are as follows: we choose three toy models with quadratic power, cubic power, and slowly decaying forms. Among these, the quadratic and cubic power models are applicable to the optically thin accretion disks observed by the EHT; the cubic power model (with $r_{\text{ph}}$ as the break point) can more accurately reflect the intensity variations of the photon ring. The specific forms are given below.

Emission model 1: Quadratic power-law decay
\begin{equation}
I_{\text{em}}(r) =
\begin{cases}
I_0 \left[ \dfrac{1}{r - (r_{\text{isco}} - 1)} \right]^2, & r > r_{\text{isco}}, \\[6pt]
0, & r \leq r_{\text{isco}}
\end{cases} \label{eq:model1}
\end{equation}

Emission model 2: Cubic power-law decay
\begin{equation}
I_{\text{em}}(r) =
\begin{cases}
I_0 \left[ \dfrac{1}{r - (r_{\text{ph}} - 1)} \right]^3, & r > r_{\text{ph}}, \\[6pt]
0, & r \leq r_{\text{ph}}
\end{cases} \label{eq:model2}
\end{equation}

Emission model 3: Slowly decaying function
\begin{equation}
I_{\text{em}}(r) =
\begin{cases}
I_0 \frac{\frac{\pi}{2} - \tan^{-1}[r - (r_{isco} - 1)]}{\frac{\pi}{2} - \tan^{-1}[r_h - (r_{isco} - 1)]}, & r > r_h \\[10pt]
0, & r \leq r_h
\end{cases} \label{eq:model3}
\end{equation}

In this context,, $r_h$ denotes the BH EH, $k_{\text{ph}}$ stands for the critical impact parameter, $r_{\text{ph}}$ represents the photon sphere radius, and $r_{\text{isco}}$ refers to the radius of the ISCO. Table~\ref{tab:1} lists the numerical data associated with these quantities.

\begin{figure*}[!t]
    \centering
    \begin{minipage}{0.45\linewidth}
        \centering
        \includegraphics[width=\linewidth]{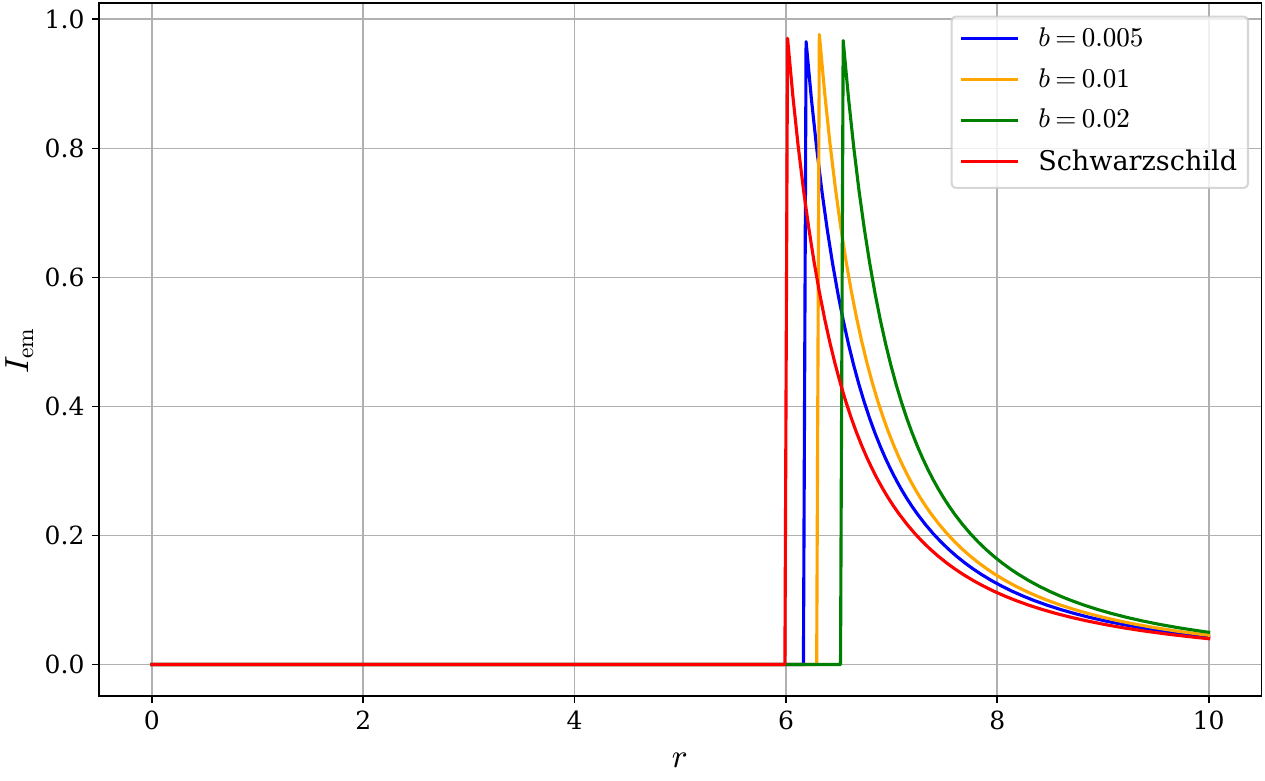}
        \textbf{(a)}
        \label{fig:7a}
    \end{minipage}
    \hfill
    \begin{minipage}{0.45\linewidth}
        \centering
        \includegraphics[width=\linewidth]{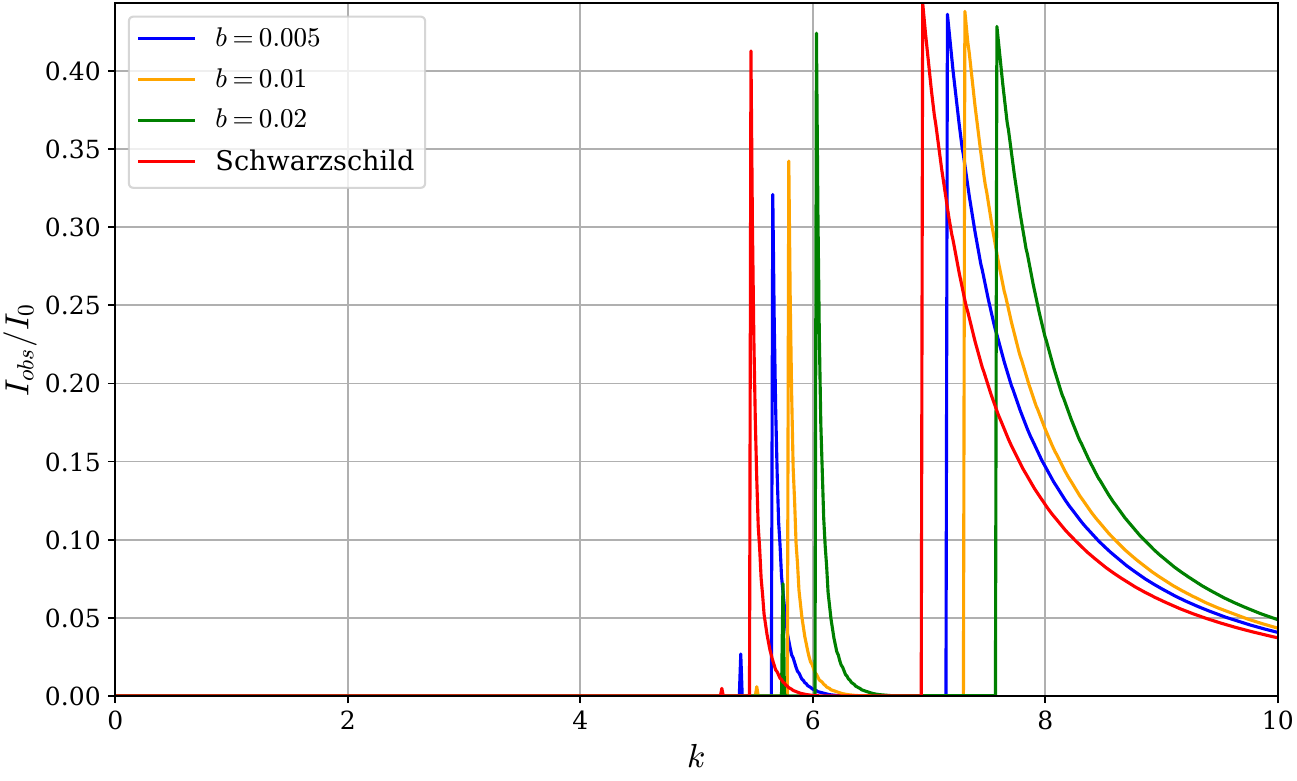}
        \textbf{(b)}
        \label{fig:7b}
    \end{minipage}
    \\[0.2cm]
    \begin{minipage}{0.220\linewidth}
        \centering
        \includegraphics[width=\linewidth]{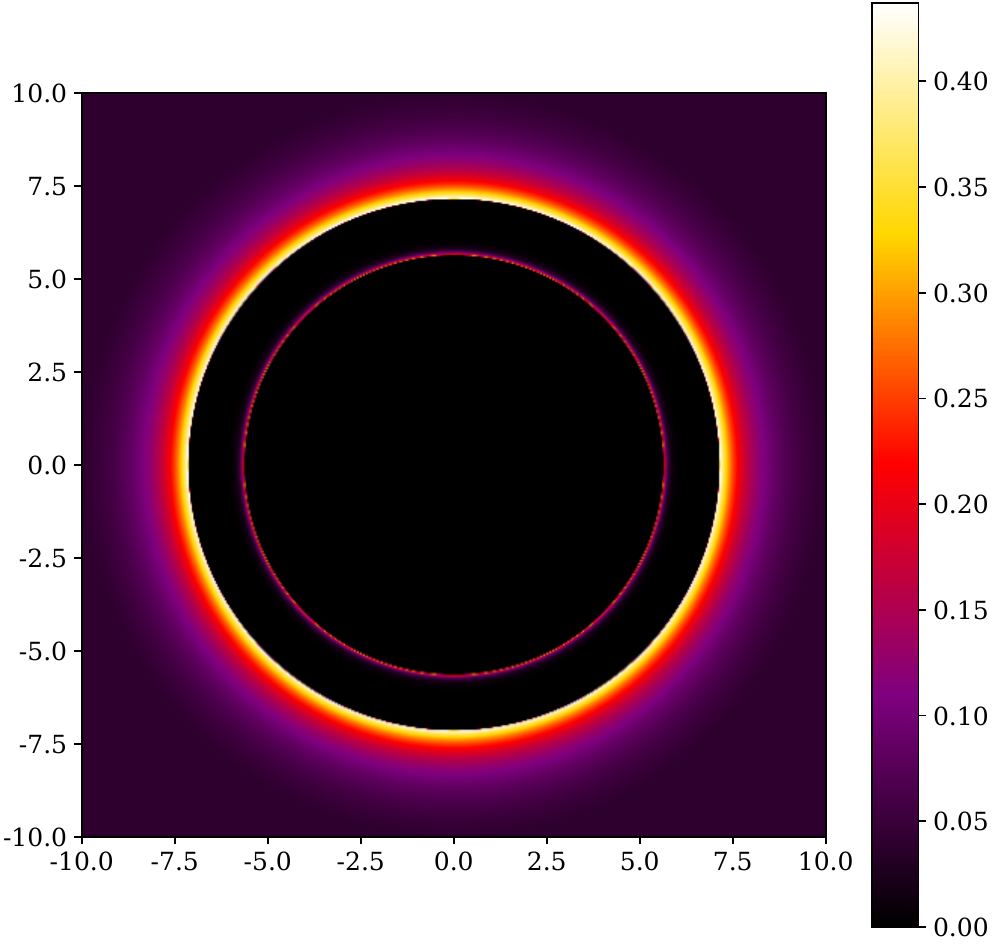}
        \textbf{(c1) $0.005$}
        \label{fig:7c1}
    \end{minipage}
    \hfill
    \begin{minipage}{0.220\linewidth}
        \centering
        \includegraphics[width=\linewidth]{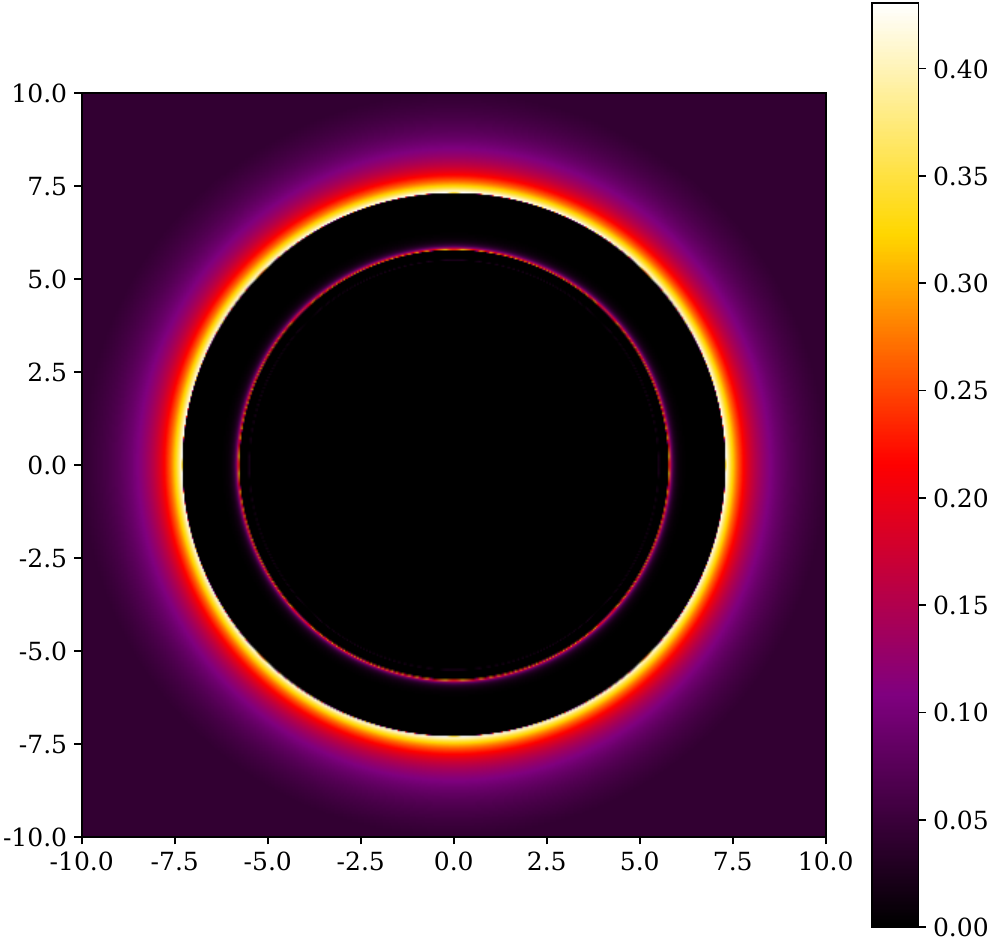}
        \textbf{(c2) $0.01$}
        \label{fig:7c2}
    \end{minipage}
    \hfill
    \begin{minipage}{0.220\linewidth}
        \centering
        \includegraphics[width=\linewidth]{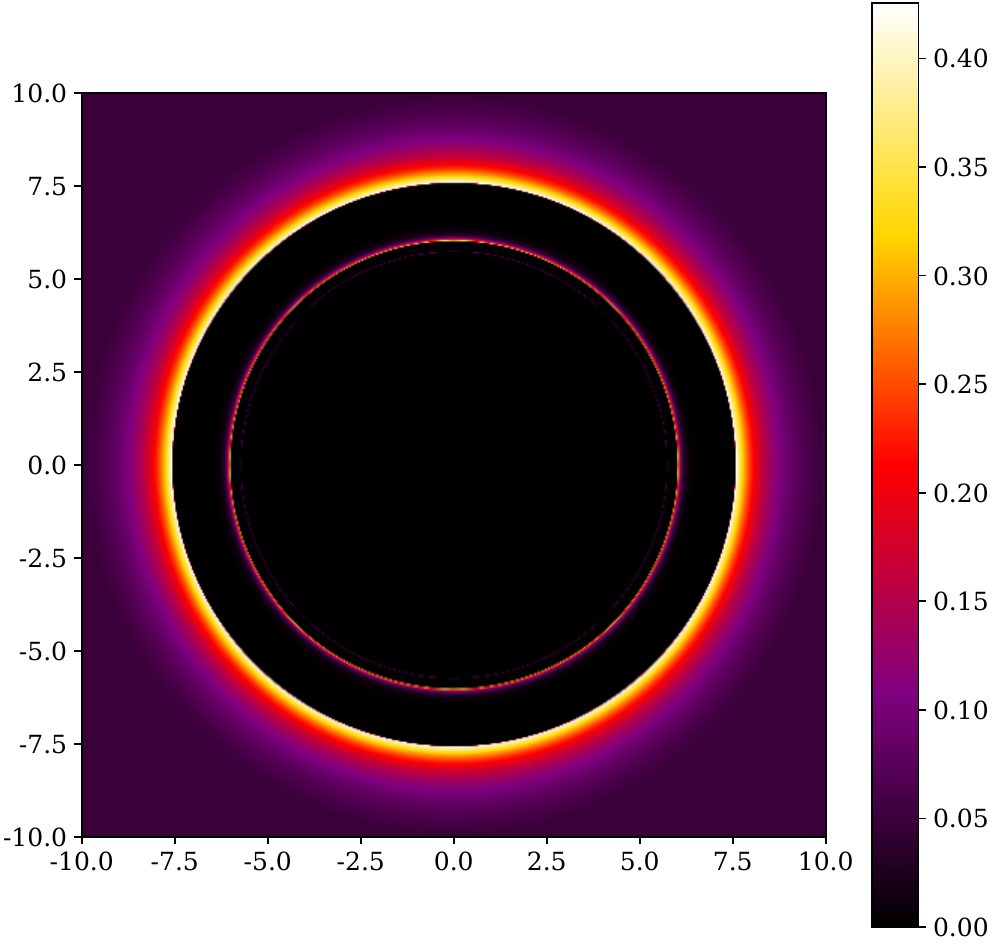}
        \textbf{(c3) $0.02$}
        \label{fig:7c3}
    \end{minipage}
    \hfill
    \begin{minipage}{0.220\linewidth}
        \centering
        \includegraphics[width=\linewidth]{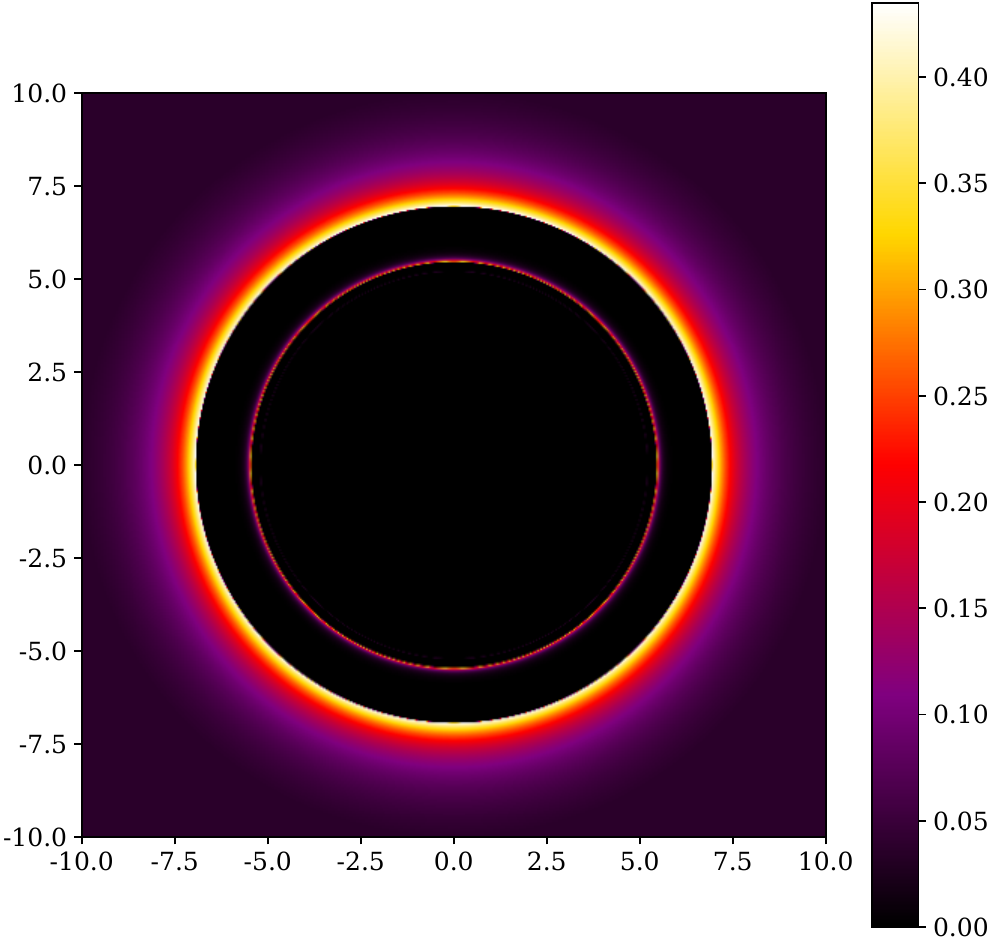}
        \textbf{(cs) Schwarzschild}
        \label{fig:7cs}
    \end{minipage}
    \caption{First emission model. (a) Emission intensity vs. radius $r$; (b) total observed intensity vs. impact parameter $k$; (c) optical appearance from the observer's perspective. Columns (left to right): $b = 0.005$, $0.01$, $0.02$, and the Schwarzschild case($P=0,b \approx 0$).}
    \label{fig7}
\end{figure*}

\begin{figure*}[!t]
    \centering
    \begin{minipage}{0.45\linewidth}
        \centering
        \includegraphics[width=\linewidth]{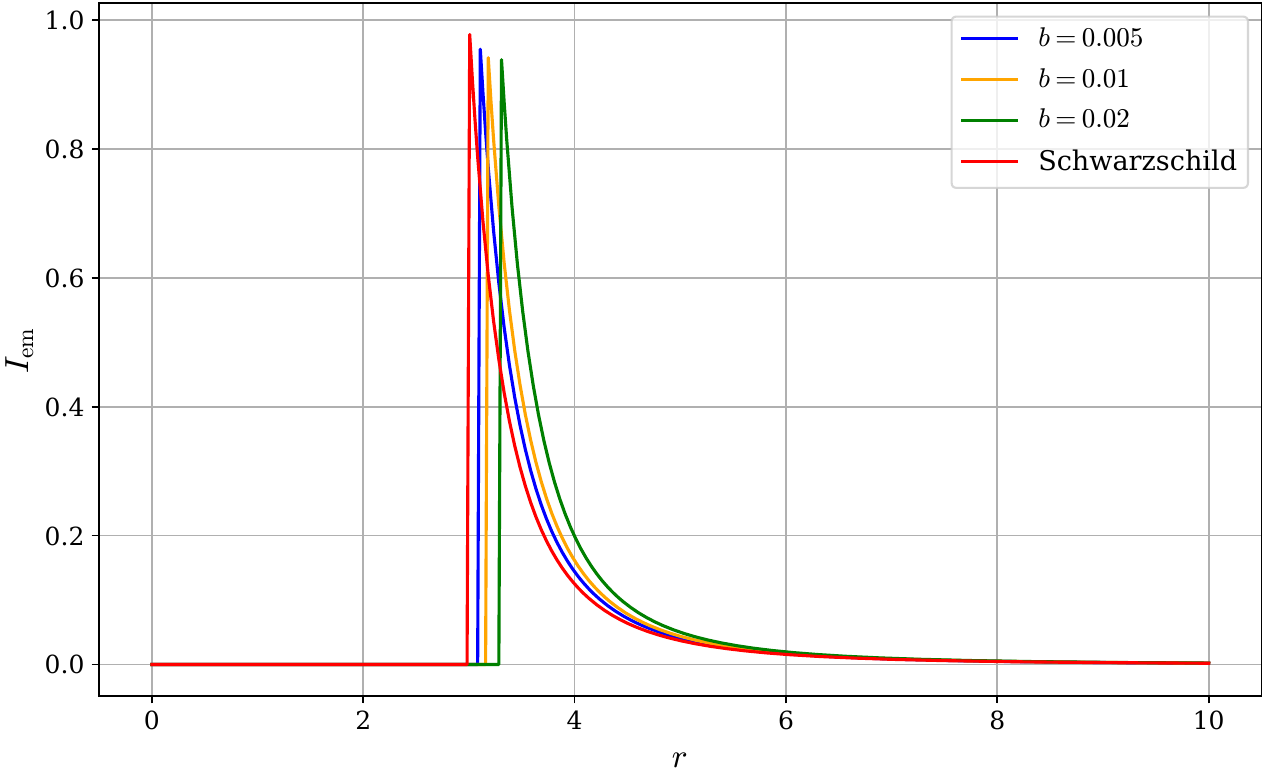}
        \textbf{(d)}
        \label{fig:8d}
    \end{minipage}
    \hfill
    \begin{minipage}{0.45\linewidth}
        \centering
        \includegraphics[width=\linewidth]{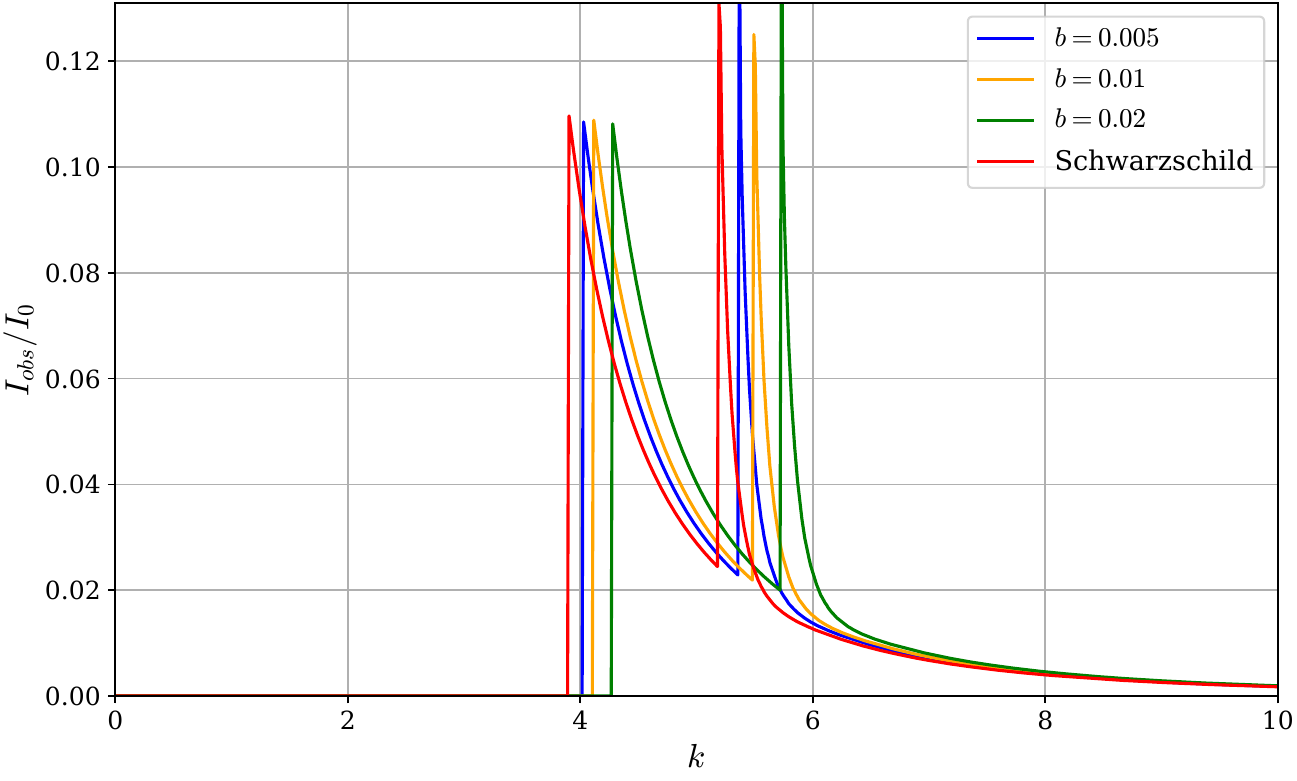}
        \textbf{(e)}
        \label{fig:8e}
    \end{minipage}
    \\[0.12cm]
    \begin{minipage}{0.220\linewidth}
        \centering
        \includegraphics[width=\linewidth]{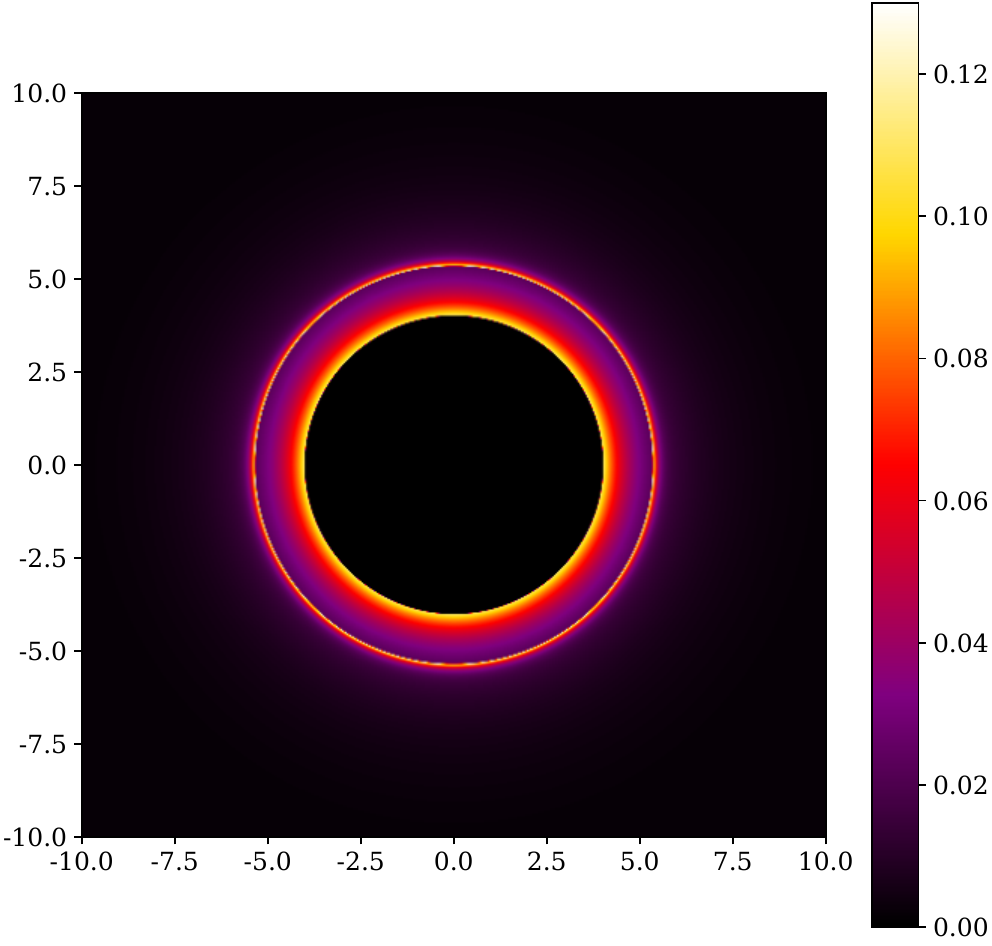}
        \textbf{(f1) $0.005$}
        \label{fig:8f1}
    \end{minipage}
    \hfill
    \begin{minipage}{0.220\linewidth}
        \centering
        \includegraphics[width=\linewidth]{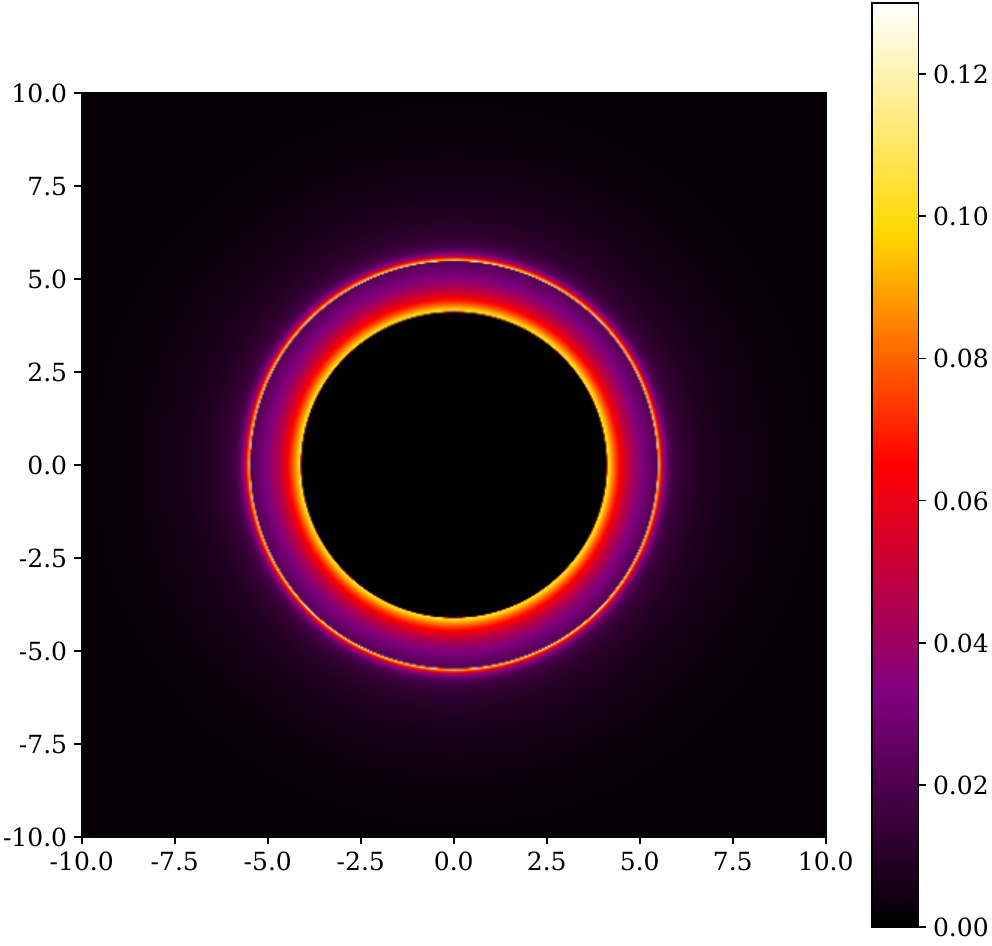}
        \textbf{(f2) $0.01$}
        \label{fig:8f2}
    \end{minipage}
    \hfill
    \begin{minipage}{0.220\linewidth}
        \centering
        \includegraphics[width=\linewidth]{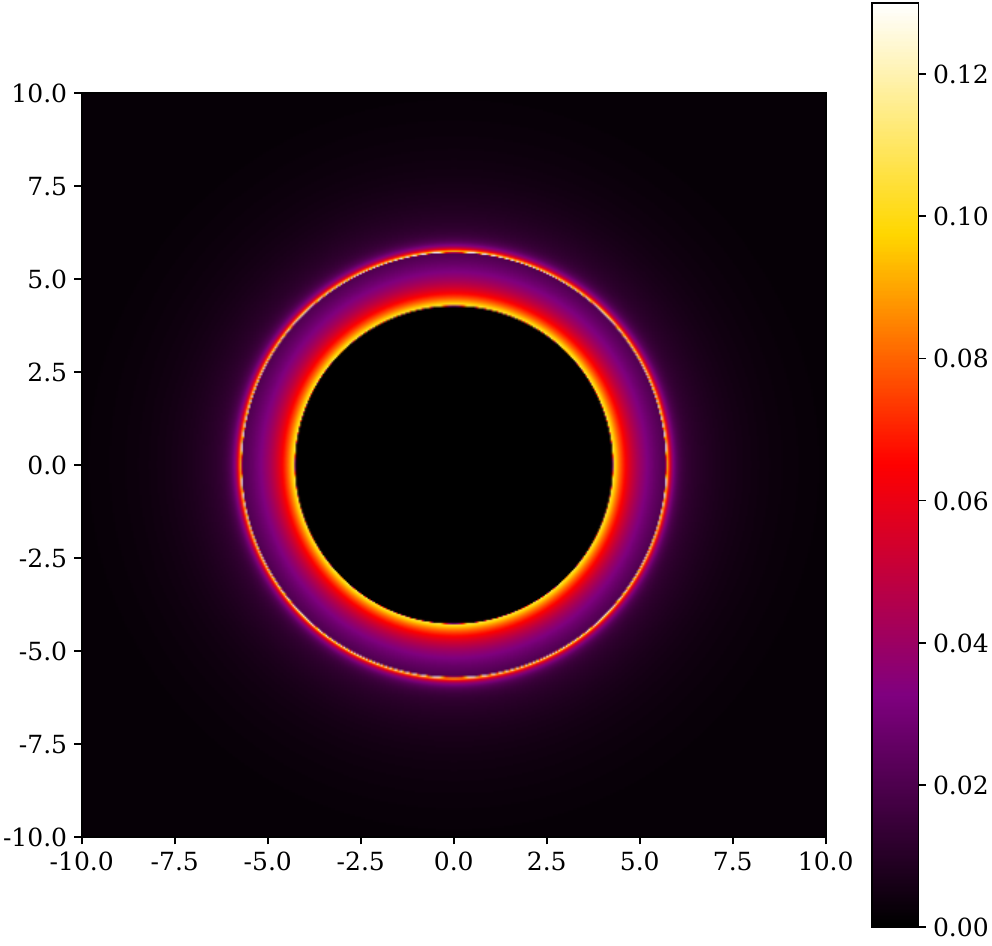}
        \textbf{(f3) $0.02$}
        \label{fig:8f3}
    \end{minipage}
    \hfill
    \begin{minipage}{0.220\linewidth}
        \centering
        \includegraphics[width=\linewidth]{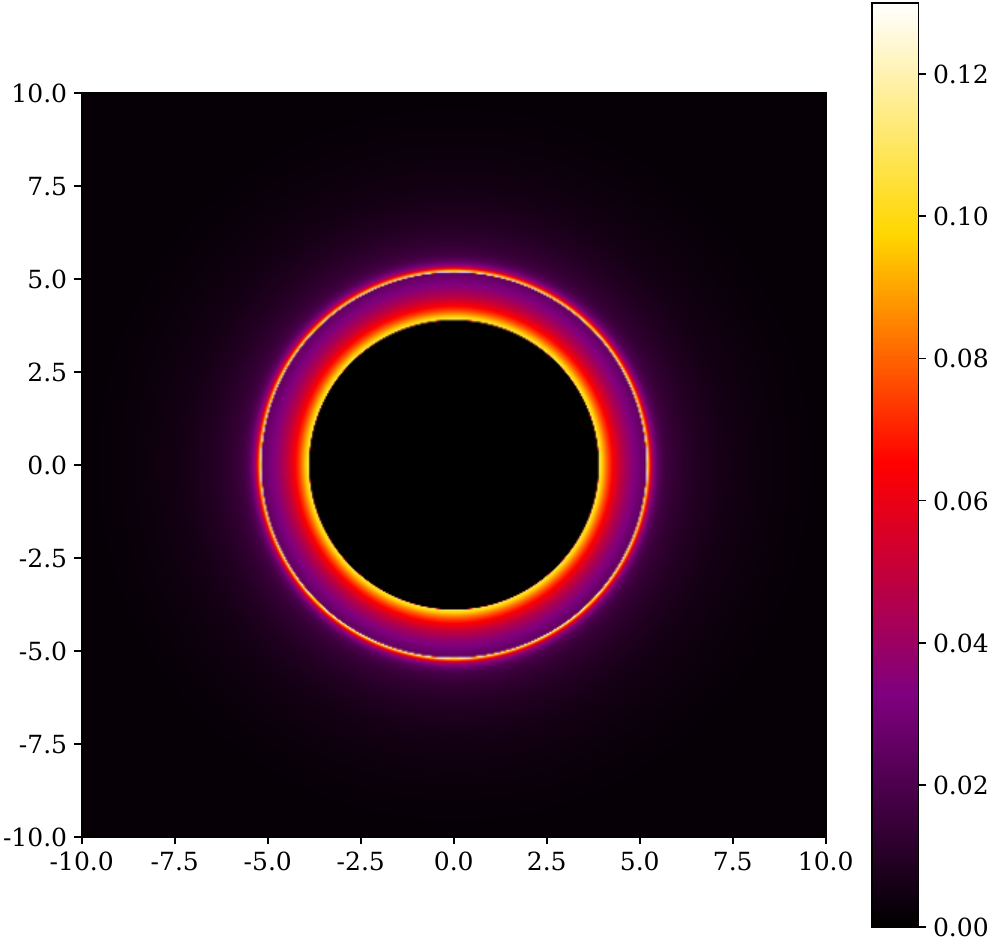}
        \textbf{(fs) Schwarzschild}
        \label{fig:8fs}
    \end{minipage}

    \caption{Second emission model. (d) Emission intensity vs. radius $r$; (e) total observed intensity vs. impact parameter $k$; (f) optical appearance from the observer's perspective. Columns (left to right): $b = 0.005$, $0.01$, $0.02$, and the Schwarzschild case($P=0,b \approx 0$).}
    \label{fig8}
\end{figure*}

\begin{figure*}[!t]
    \centering
    \begin{minipage}{0.45\linewidth}
        \centering
        \includegraphics[width=\linewidth]{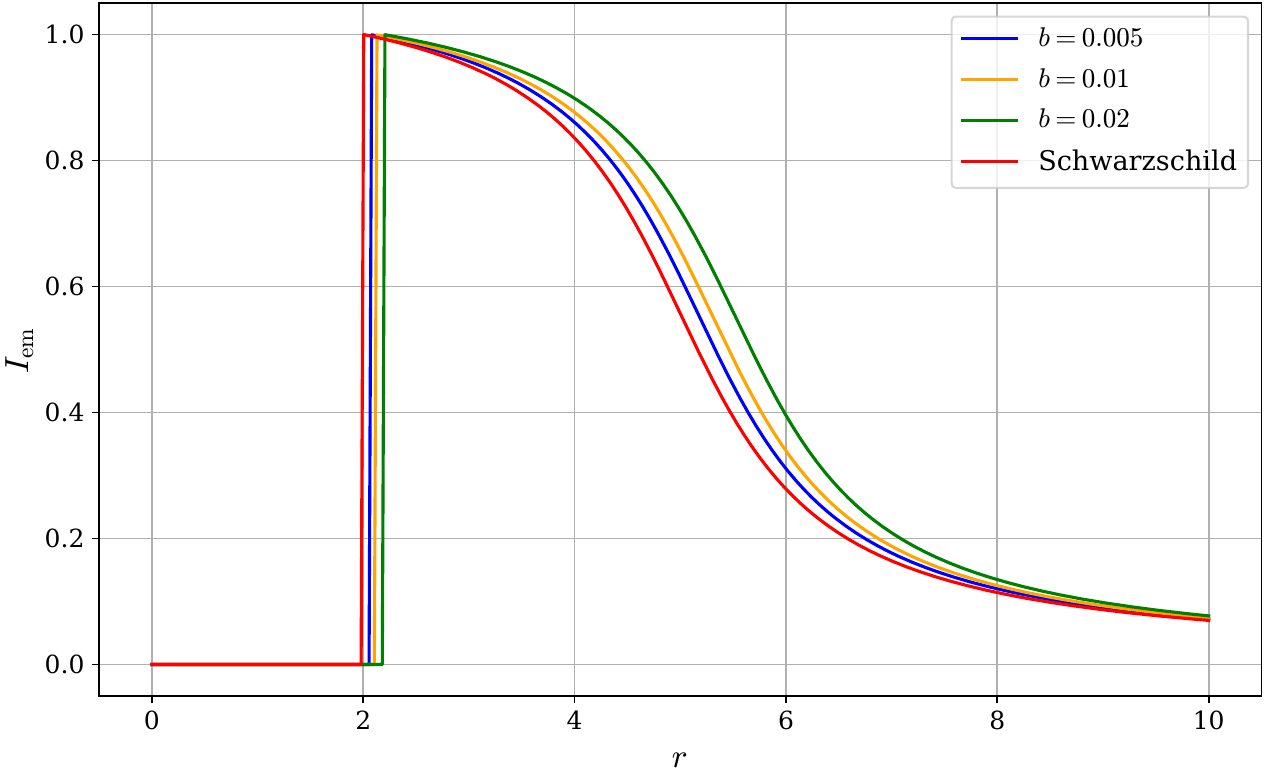}
        \textbf{(g)}
        \label{fig:9g}
    \end{minipage}
    \hfill
    \begin{minipage}{0.45\linewidth}
        \centering
        \includegraphics[width=\linewidth]{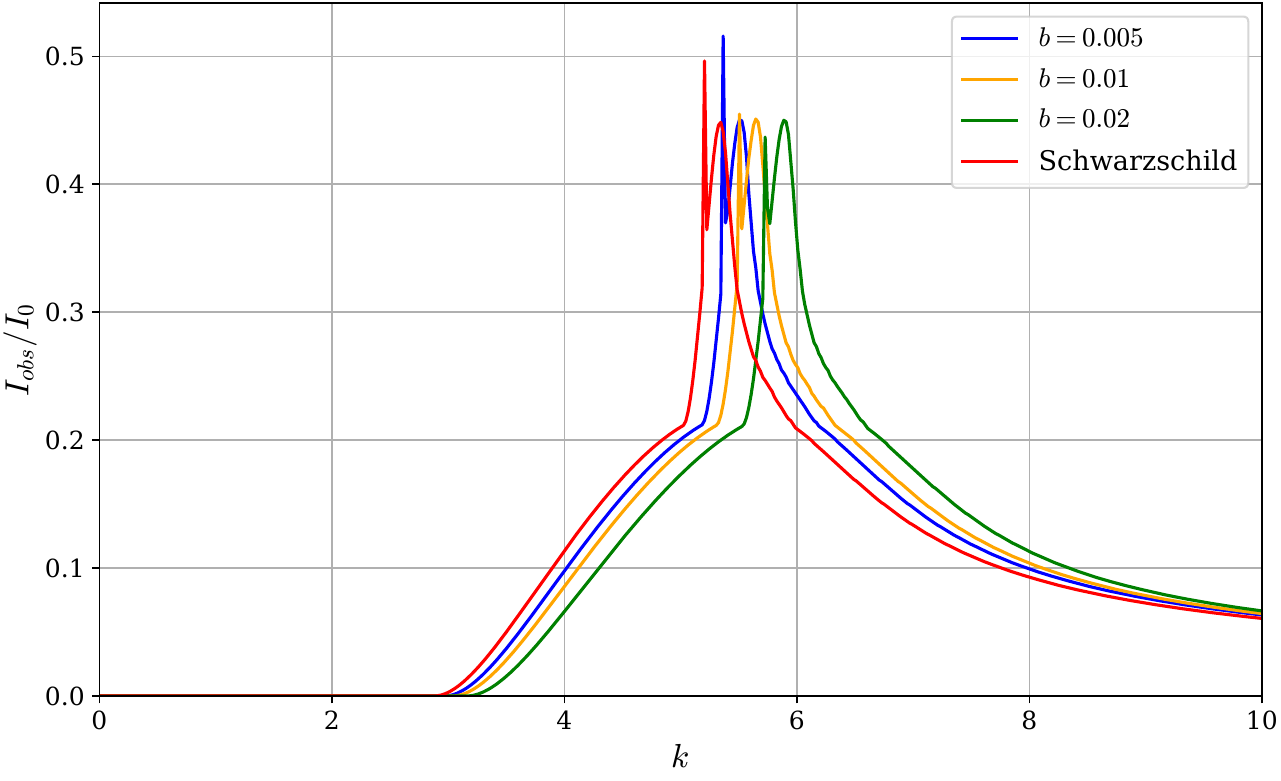}
        \textbf{(h)}
        \label{fig:9h}
    \end{minipage}
    \\[0.12cm]
    \begin{minipage}{0.220\linewidth}
        \centering
        \includegraphics[width=\linewidth]{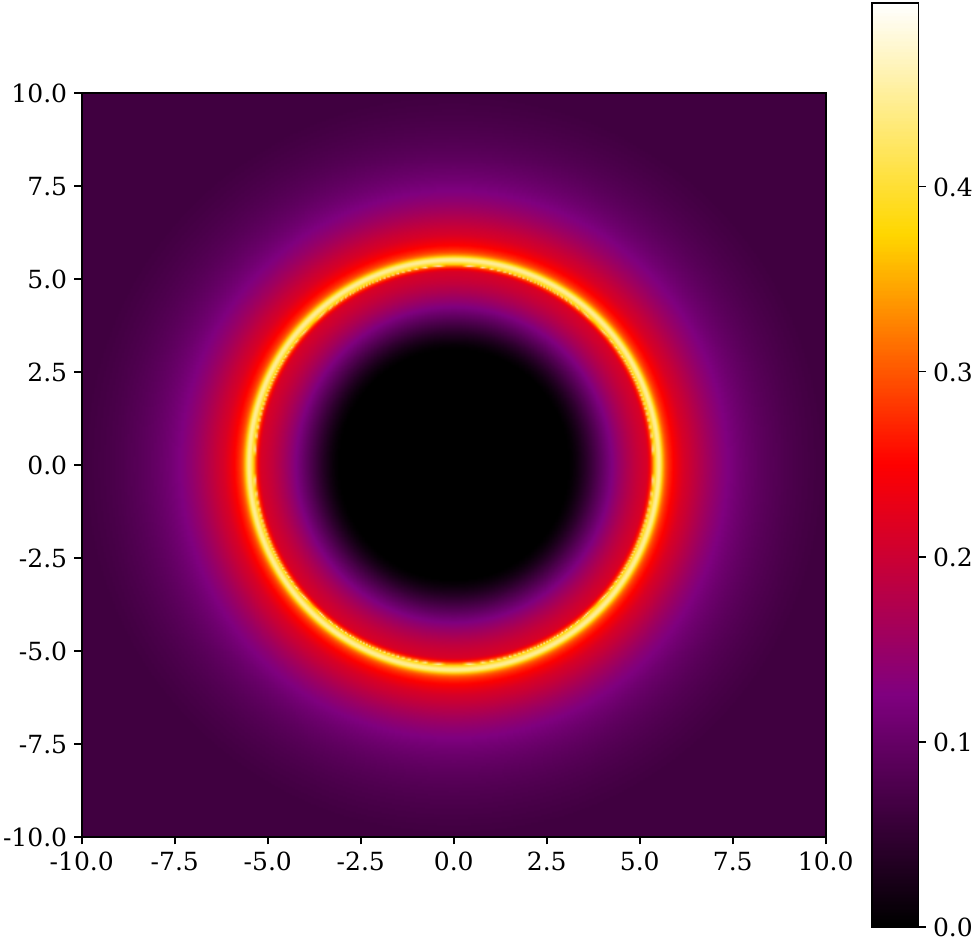}
        \textbf{(i1) $0.005$}
        \label{fig:9i1}
    \end{minipage}
    \hfill
    \begin{minipage}{0.220\linewidth}
        \centering
        \includegraphics[width=\linewidth]{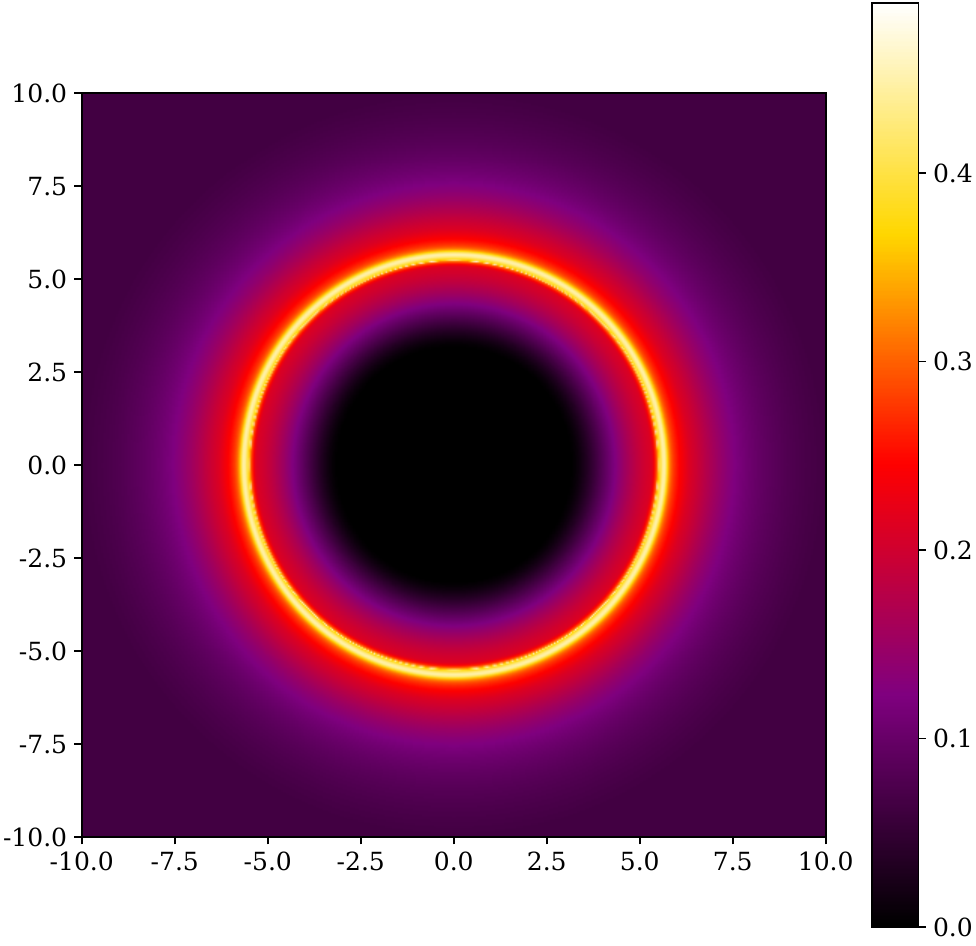}
        \textbf{(i2) $0.01$}
        \label{fig:9i2}
    \end{minipage}
    \hfill
    \begin{minipage}{0.220\linewidth}
        \centering
        \includegraphics[width=\linewidth]{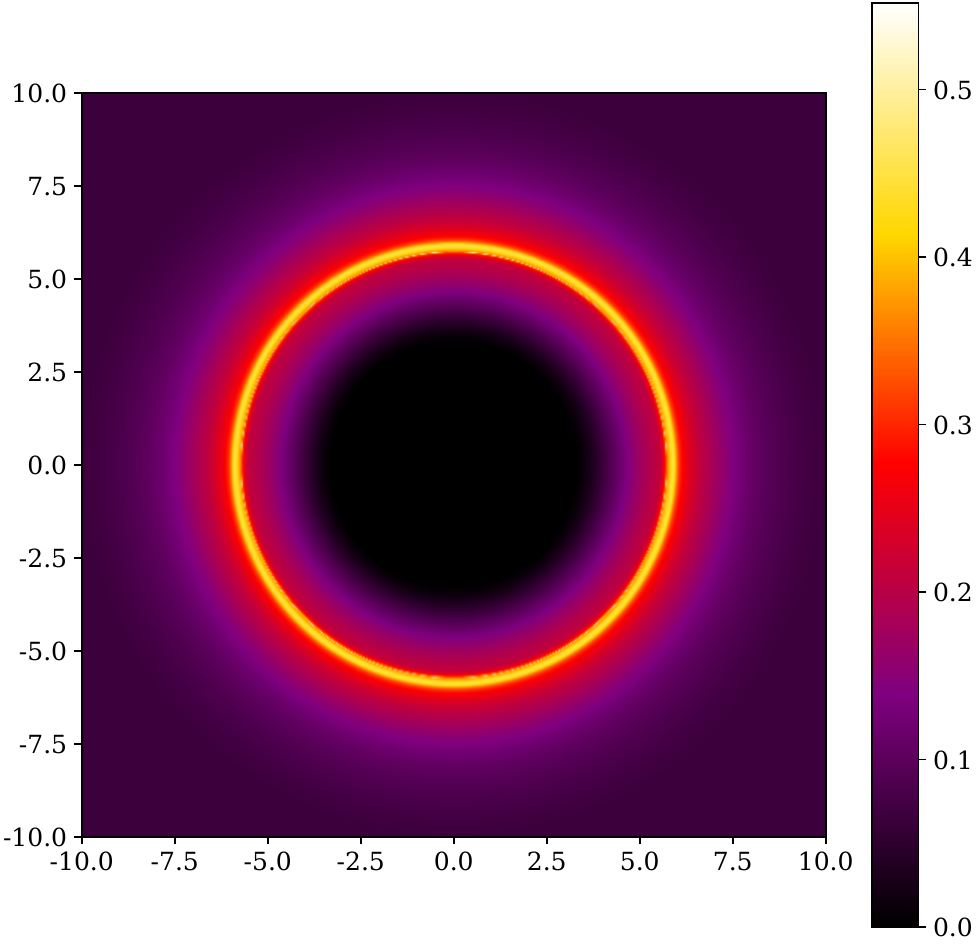}
        \textbf{(i3) $0.02$}
        \label{fig:9i3}
    \end{minipage}
    \hfill
    \begin{minipage}{0.220\linewidth}
        \centering
        \includegraphics[width=\linewidth]{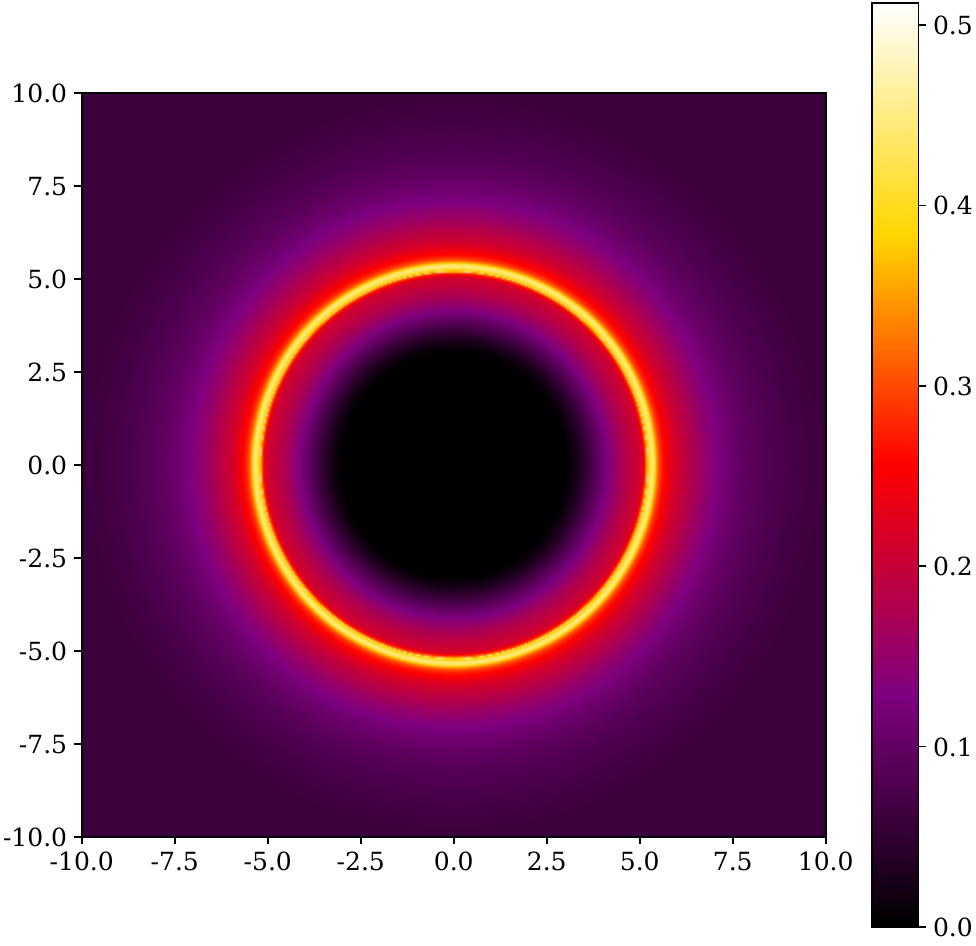}
        \textbf{(is) Schwarzschild}
        \label{fig:9is}
    \end{minipage}

    \caption{Third emission model. (g) Emission intensity vs. radius $r$; (h) total observed intensity vs. impact parameter $k$; (i) optical appearance from the observer's perspective. Columns (left to right): $b = 0.005$, $0.01$, $0.02$, and the Schwarzschild case($P=0,b \approx 0$).}
    \label{fig9}
\end{figure*}

Through combining the emission intensity functions and transfer functions with the total observed intensity formula Eq.~\eqref{eq:36}, and applying the spherically symmetric projection of the observed intensity, we generate the two-dimensional optical appearance images for each set of parameters, as shown in Figs.~\ref{fig7}, \ref{fig8}, and \ref{fig9}.
\renewcommand{\dblfloatpagefraction}{1}
\renewcommand{\dbltopfraction}{1}

In Figs.~\ref{fig7}, \ref{fig8}, and \ref{fig9}, the quadratic power-law decay model is illustrated in panels (a), (b), and (c); the cubic power-law decay model is shown in panels (d), (e), and (f); and the slowly decaying model is presented in panels (g), (h), and (i). Within panels (c), (f), and (i), the subscripts 1, 2, 3, and s are employed to denote the optical appearances of the accretion disk corresponding to $b = 0.005$, $b = 0.01$, $b = 0.02$, and the Schwarzschild case($P=0,b \approx 0$), respectively.

For the three toy models, the emission intensity is plotted against the radius $r$ in panels (a), (d), and (g), and the total observed intensity against the $k$ in panels (b), (e), and (h); and the optical appearance perceived by the observer for various values of the BH molecular volume parameter $b$ under the three toy models is illustrated in panels (c), (f), and (i).

Figs.~\ref{fig7}, \ref{fig8}, and \ref{fig9} clearly illustrate the optical appearance characteristics of the Van der Waals BH for different values of $b$ and different emission models:
\begin{itemize}
    \item For the quadratic power-law decay model, as shown in Figs.~\ref{fig7}(a,b,c1,c2,c3,cs), when $b = 0.005$ (Fig.~\ref{fig7}(c1)), the photon ring exhibits relatively concentrated brightness and a well-defined profile, with only a slight difference from the Schwarzschild BH photon ring (Fig.~\ref{fig7}(cs)). As $b$ increases to $0.01$ (Fig.~\ref{fig7}(c2)) and then to $0.02$ (Fig.~\ref{fig7}(c3)), the photon ring radius gradually enlarges, while the brightness peak decreases slightly but remains concentrated. Such behavior agrees with the tendency for the photon sphere radius $r_{\text{ph}}$ to grow as $b$ becomes larger, as tabulated in Table~\ref{tab:1}.

    \item For the cubic power-law decay model, as shown in Figs.~\ref{fig8}(d,e,f1,f2,f3,fs), this model provides a more accurate depiction of the photon ring details and enhances the contrast between the photon ring and the accretion disk brightness. Correspondingly, in the observed intensity function curve (Fig.~\ref{fig8}(e), two widely separated peaks can be clearly distinguished. After mapping to the image plane, a very distinct double-ring structure is observed in the optical appearance images (Fig.~\ref{fig8}(f)).

    \item For the slowly decaying model, as shown in Figs.~\ref{fig9}(g,h,i1,i2,i3,is), although the radiation intensity decays slowly, the photon rings corresponding to different values of $b$ are readily distinguishable. This is because a larger $b$ leads to a more spread-out photon ring, which is physically consistent with the mechanism that an increase in the BH molecular volume modifies the spacetime curvature distribution and shifts the photon orbits outward.
\end{itemize}

A contrast between the optical appearance of the Van der Waals BH and that of the Schwarzschild BH (as illustrated in the subfigures marked with the subscript $s$ in Figs.~\ref{fig7}, \ref{fig8}, and \ref{fig9})reveals that the photon ring of the former responds with pronounced sensitivity to changes in $b$, while the latter displays no dependence on this parameter whatsoever. Such a discrepancy, stemming from the SBH--LBH phase transition together with the microstructural discontinuities peculiar to the Van der Waals BH, furnishes a crucial diagnostic criterion for differentiating these two classes of BHs via observational data.

The change in the optical appearance of the Van der Waals BH arises fundamentally as the macroscopic consequence of the manner in which the molecular volume parameter $b$ governs the microscopic number density $n$: as $b$grows, the number density $n$ declines, and the weak attractive interactions among the BH molecules become diminished ~\cite{ref5}, which means that the geometry of spacetime is no longer the same in the region close to the BH horizon, ultimately appearing as an enlargement of the photon ring radius together with a modest reduction in the brightness peak ~\cite{ref13}.

\section{CONCLUSIONS AND DISCUSSION}
\label{sec:level5}
The new BH images released by the EHT collaboration~\cite{ref18,ref73} provide unprecedented observational support for testing non-Kerr BH models using photon rings~\cite{ref42,ref68}. Meanwhile, these developments also offer novel pathways for following null geodesics in strong gravitational fields, assessing general relativity’s predictions in extreme regimes, and uncovering the spacetime geometry in the vicinity of BHs ~\cite{ref74,ref75,ref76}. To this end, we select the Van der Waals BH, which is characterized by thermodynamic phase transitions, as the research background of this paper. Under the intensity emitted by a given toy accretion model, we employ the photon ring to test the BH molecular model, obtain the optical appearances of the Van der Waals BH under three toy accretion models, and investigate the influence of the BH molecular volume parameter $b$ on its image structure.

In this study, we first fixed the parameter $a$ in the Van der Waals BH metric, which characterizes the attractive interaction between BH molecules and the thermodynamic pressure $P$, while keeping the remaining parameters fixed, so that the effect of the molecular volume parameter $b$ could be examined independently. As the molecular volume parameter $b$ varies, the EH of the Van der Waals BH shifts only modestly; in the limiting case $b$ approaches zero, the metric goes over to the Schwarzschild form. We then employ the Lagrangian formalism to obtain the null geodesic equations for photons together with the corresponding constants of motion. These enable an efficient determination of the requirements that must be satisfied for the photon sphere to exist, along with the associated radii and parameter values. In addition, the shadow measurements of M87* and Sgr A* reported by the EHT collaboration are used to place bounds on the molecular volume parameter $b$ of the BH. At the same time, three emission models are adopted to simulate the BH's optical appearance, and the influence of $b$ on the photon ring structure and brightness is examined quantitatively.

It is observed that, for the static spherically symmetric Van der Waals BH, a positive correlation is exhibited between the BH molecular volume parameter $b$ and each of the following quantities: $r_h$, $r_{\text{ph}}$,  $k_{\text{ph}}$, and  $r_{\text{isco}}$. When a static thin accretion disk is present, these quantities help determine the null geodesic paths, the shadow, and what the photon ring looks like optically. Moreover, the molecular volume parameter $b$ is found to strongly affect the properties of the photon ring. For $b \in [0.005, 0.02]$, the photon sphere radius increases from 3.0923 to 3.2870, indicating that the photon ring radius expands synchronously with increasing $b$. For M87*, this interval shows good agreement with observations; for Sgr A*, only the subrange $b \in [0, 0.008]$ is consistent with observations, while still satisfying the energy conditions and the requirements of stable phase transitions. These findings indicate that the photon ring of the BH provides an effective probe of the physical parameters characterizing the Van der Waals BH model.

The above phenomena are not coincidental---they are underpinned by a clear microscopic physical mechanism. Through thermodynamic correlation analyses, we further reveal the microscopic origin of the molecular volume parameter $b$ in regulating photon ring properties: the molecular volume parameter $b$ is the macroscopic equivalent of the microscopic molecular volume of the Van der Waals BH, and is negatively correlated with the BH molecular number density $n$ ($n =1/(2r_h+3b)$). More specifically, an increase in $b$ reduces the number density $n$, thereby suppressing the weak attractive interactions between BH molecules. This changes the gradient of the spacetime curvature, displaces the photon sphere radius outward, and eventually modifies both the structure and the brightness of the photon ring. This connection links the microscopic thermodynamic properties (number density $n$), the macroscopic spacetime geometry (metric and horizon radius $r_h$), and the observable photon ring characteristics (radius, brightness, and structure) into a complete chain, providing a new pathway for inferring the microscopic physical properties of BHs from observational data.

Quantitative results show that for $b \in [0.005, 0.02]$, the number density $n$ decreases from 0.2418 to 0.2254 as $b$ increases, while the corresponding photon sphere radius increases from 3.0923 to 3.2870. For M87*, this interval gives the best agreement with observations; for Sgr A*, the best-agreement subrange is $b \in [0, 0.008]$. Both subranges simultaneously satisfy the energy conditions and stable phase transition requirements, further validating the physical consistency of the microscopic number density correlation.

It is worth noting that we have not taken into account the effects of rotation and charge, whereas in realistic astrophysical BHs, most are not static but possess angular momentum and electric charge, and their optical appearance is significantly influenced by spin and accretion disk type~\cite{ref38}. Moreover, the presence of dark matter halos can modify the observed images of BHs ~\cite{ref77}, while geodesic motion may encode signatures of new physics, such as neutrino propagation ~\cite{ref78}. Furthermore, BHs in Horndeski gravity also exhibit distinctive optical features~\cite{ref45}. Our consideration solely from the static spherically symmetric perspective may affect the practical applicability of the model. Moreover, the present analysis has focused solely on the effect of the BH molecular volume parameter in the Van der Waals BH metric.

In future work, we will further incorporate the inter-molecular interaction parameter $a$ and the pressure parameter $P$ of the fluid BH, systematically investigate the coupling effects among multiple parameters and their joint influence on observables, and employ parameter estimation methods such as Fisher matrix or Bayesian analysis to characterize the correlations and degeneracies among parameters through corner plots. These leave numerous directions worthy of in-depth exploration for subsequent research. For example, one could generalize the analysis to the rotating Van der Waals BH model and, by adapting the optical-appearance techniques developed for rotating hairy BHs, examine the joint effects of spin and the molecular volume parameter $b$ on the photon ring~\cite{ref38}. It is also possible to combine gravitational wave data to construct a photon ring--gravitational wave multimessenger constraint framework~\cite{ref64}. Meanwhile, close attention should be paid to the parameter degeneracy issue~\cite{ref79}, and the higher-order structure of the photon ring can be utilized to further enhance model discriminability~\cite{ref40}. Moreover, the photon-ring properties of regular BHs in asymptotically safe gravity ~\cite{ref46}, as well as those of scalar hairy BHs with an inverted Higgs potential ~\cite{ref14}, provide useful benchmarks for future comparative studies.

Overall, through theoretical derivation and observational simulations, this study reveals the potential role of BH photon rings in distinguishing between different BH models, and confirms the feasibility of using photon rings to test Van der Waals BHs. It is hoped that the present work may pave the way for further explorations of broader spacetime structures and quantum gravity theories, and that it may also stimulate new insights and more dedicated efforts in the study of BH shadows and other hairy BH models.

\begin{acknowledgments}
This work was supported by National Natural Science Foundation of China (Grant No.12365008), the Guizhou Provincial Basic Research Program (Natural Science) (Grant No.QianKeHe-JiChu[2024]Young166), the Guizhou Provincial Basic Research Program (Natural Science) (Grant No.QianKeHeJiChu-ZK[2024] YiBan027 and QianKeHeJiChuMS[2025]680), the Guizhou Provincial Major Scientific and Technological Program XKBF(2025) 010(Hosted by Professor Xu Ning), the Guizhou Provincial Major Science and Technological Program XKGF (2025) 009 (Hosted by Professor Xiang Guoyong) and Guizhou Provincial Major Scientific and technological Program (Teacher Fan Lu Lu moderated).  
\end{acknowledgments}
\nocite{*}
\bibliographystyle{unsrt}
\bibliography{wenxian}
\end{document}